\documentclass[aps,nofootinbib,twocolumn]{revtex4}
\usepackage{graphicx,amsmath,hyperref,amssymb,bbm,expdlist}

\newcommand{\be}{\begin{equation}}
\newcommand{\ee}{\end{equation}}
\def\C{{\mathbbm C}}

\def\C{{\cal C}}

\begin{document}
\title{ \Large Loop quantum gravity: the first twenty five years}
     \author{Carlo Rovelli}
     \affiliation{Centre de Physique Th\'eorique de Luminy\footnote{Unit\'e mixte de recherche du CNRS et des Universit\'es de Aix-Marseille I, Aix-Marseille II et Toulon-Var; affili\'e \`a la FRUMAM.}, Case 907, F-13288 Marseille, EU}
\date{\small  \today}
\begin{abstract}
\noindent 
I give a synthetic presentation of loop quantum gravity.  I spell-out the aims of the theory and compare the results obtained with the initial hopes that motivated the early interest in this research direction. I give my own perspective on the status of the program and attempt of a critical evaluation of its successes and limits.
\end{abstract}
\maketitle

\section{Introduction}

Loop gravity is not quite twenty-five years old, but is getting close to such a venerable age: several basic ideas emerged from extensive discussions at
a 1986 workshop on Quantum Gravity at the Institute of Theoretical Physics in Santa Barbara, and the first conference talk on a ``loop-space representation of quantum general relativity" was given at a conference in Goa, India, in 1987 \cite{RovelliSmolin87}. The theory have been much growing since, with the contribution of a considerable number of researchers. It has raised hopes, generated fascination, encounter fierce opposition, and is today developed in several directions by about forty research groups around the world.  Responding to the invitation by ``Classical and Quantum Gravity", I try here to assess what the theory has produced so far, the extent to which the initial hopes have been realized, and the  disappointments, surprises and successes that the theory has encountered. 

It is a particularly appropriate moment for such an assessment because the last few years have been especially productive in several directions, and have substantially modified the the lay of the land. This is therefore a good opportunity for taking stock of loop gravity.   

In spite of repeated efforts by its senior members (including, alas, myself) the loop research community has been characterized since its beginning by a multi-faced, almost anarchic, internal structure, where different individuals hold strong and often conflicting views. In spite of it this, the theory has evolved very coherently.  The plurality of points of views has generated confusion, and still does, but also a lively dynamics that has lead to richness and results.  Here I definitely do not attempt to give a complete overview of the landscape of opinions and perspectives on the theory. I give my own view and understanding of the state of the art. Also, this note is far from being comprehensive, and I apologize for the topics I do not cover.  In particular, this review is centered on the covariant theory, a bit disregarding the the Hamiltonian picture \cite{ThiemannBook,Giesel:2009jp,Domagala:2010bm}.  Various perspectives on the theory can be found in several books \cite{Rovelli:2004fk,ThiemannBook,BaezBook2,Gambini,Oriti:2007kx} and review or introduction articles \cite{Ashtekar:2004eh,Ashtekar:2007px,Ashtekar:2007tv,Smolin:2004sx,Leecourse,Freidel:2005fk,Rovelli:2010wq,Rovelli:2010vv,Seth}.  

The history of quantum gravity is full of great hopes later disappointed.  I remember as a young student sitting in a major conference where a world-renowned physicists announced that the final theory of quantum gravity and everything had finally been found.  There were a few skeptics in the audience, regarded as stupid zombies by the large majority. Today most of us do not even remember the name of that ``final theory", or, worse, can think of more than one possibility...  I hope that for a problem as difficult as quantum gravity we all keep in mind the sequence of failures behind us, and be very aware that what we are doing, even if we are many and all convinced, could nevertheless be factually wrong.  

But I want also to try to counterbalance any a priori pessimism with the consideration that major problems of this kind have sometimes resisted for a while in the history of physics, but then a solution has generally been found. My own final evaluation is that loop gravity might turn out to be wrong as a theoretical description of quantum spacetime, but it might also well turn out to be substantially correct. I am well aware that the issue is open, but I would (and do) bet in its favor. Internal consistency, full agreement with known low-energy physics, simplicity, and, ultimately, experience, will tell. 

I start (chapter \ref{art}) with a compact presentation of the theory, as I understand it today, in one of its possible formulations. This gives a concrete definition of the object we are talking about. In chapter \ref{problem} I discuss what are the problems that the theory intends to address. This would comes logically first, but I prefer to start concretely. I then present a historical chapter \ref{history}, reviewing the way the theory has developed, and describing the hopes, the failures and the surprises along the development of the theory. This serves also as a ``derivation" from classical general relativity of the theory given a-prioristically in chapter \ref{art}.  Chapter \ref{applications} is about the main applications and results of the theory. I attempt the assessment of the extent to which the problems are solved and the hopes achieved, or not achieved, in chapter \ref{assessment}.

\section{The theory}\label{art}

I think that a credible physical theory must be such that we can display it in a few formulas. I therefore begin with a presentation of loop gravity (in one of its possible formulations and variants) in three equations. This is going to be abstract: symbols that feature in these equations have connotations that are not at all clear at first.  I beg the reader to keep with me: the rest of this chapter is devoted to clarifying and explaining these three equations.

Quantum states of the geometry are described by functions $\psi(h_l)$ of elements $h_l\in SU(2)$ associated to the links $l$ of an arbitrary graph $\Gamma$. Transition amplitudes between such states are defined perturbatively%
\footnote{This is \emph{not} the standard quantum field theory perturbation expansion around flat space; see later.}
by  
\be
Z_{\cal C}(h_l)=   \int_{SU(2)} dh_{vf}\ \prod_f\delta(h_f)\ \prod_v A_v(h_{vf}).
\label{int1} 
\ee
$\cal C$ is a two-complex (a combinatorial set of faces $f$ that join along edges $e$, which in turn join on vertices $v$) bounded by $\Gamma$; $h_f=\prod_{v\in f} h_{vf}$ is the oriented product of the group elements around the face $f$ and the vertex amplitude is
\be
A_{v}(h_l)=\int_{SL(2,\mathbb{C})} dg'_e\ \prod_l K(h_l,g_{s_l}g^{-1}_{t_l})
\label{va}
\ee
where $s_l$ and $t_l$ are the source and target of the link $l$ in the graph $\Gamma_v$ that bounds the vertex $v$ within the two-complex. The prime on $dg_e$ indicates that one of the edge integrals is dropped (it is redundant). Finally, the kernel $K$ is 
\be
K(h,g)=\sum_j\int_{SU(2)} dk\ d_j\ \chi^j(hk) \ \chi^{\gamma j,j}(kg).
\label{K}
\ee
where $d_j=2j+1$, $\chi^j(h)$ is the spin-$j$ character of $SU(2)$ and $\chi^{\rho,n}(g)$ is the 
character of $SL(2,\mathbb{C})$ in the ($\rho,n$) unitary representation. $\gamma$ is a dimensionless parameter that characterizes the quantum theory.   

This is the theory. It is Lorentz invariant  \cite{Rovelli:2010ed}. It can be coupled to fermions and Yang-Mills fields \cite{Bianchi:2010bn,Han:2011as}, and to a cosmological constant \cite{Fairbairn:2010cp,Han:2010pz}, but I will not enter into this here. The conjecture is that this mathematics describes the quantum properties of spacetime, reduces to the Einstein equation in the classical limit, and has no ultraviolet divergences.  I now explain more in detail what the above means.

\subsection{Quantum geometry: quanta of space}\label{spin}

A key step in constructing any interactive quantum field theory is always a finite truncation of the dynamical degrees of freedom. In weakly coupled theories, such as low-energy QED or high-energy QCD, we rely on the particle structure of the free field and consider virtual processes involving finitely many, say $N$, particles, described by Feynman diagrams. These processes involve only the Hilbert space ${H}_N=\oplus_{n=1,N}H_n$, where $H_n$ is the $n$-particle state space. 
 In strongly coupled theories, such as confining QCD, we resort to a non-perturbative truncation, such as a finite lattice approximation. In both cases (the relevant effect of the remaining degrees of freedom can be subsumed under a dressing of the coupling constants and) the full theory is formally defined by a limit where all the degrees of freedom are recovered. 
 
The Hilbert space of loop gravity is constructed in a way that shares features with both these strategies. The analog of the state space ${H}_N$ in loop gravity is the space 
\be
         H_\Gamma=L_2[SU(2)^L/SU(2)^N]. 
\ee
where the states $\psi(h_l)$ live.  $\Gamma$ is an abstract (combinatorial) graph, namely a collection of links $l$ and nodes $n$, and a ``boundary" relation that associates an ordered couple of nodes $(s_l,t_l)$ (called source and target), to each link $l$.  (See the left panel in Figure 1.)  $L$ is the number of links in the graph, $N$ the number of nodes, and the $L_2$ structure is the one defined by the the Haar measure. The denominator means that the states in  $H_\Gamma$ are invariant under the local $SU(2)$ gauge transformation on the nodes
\be
  \psi(U_l)\to \psi(V_{s_l}U_lV^{-1}_{t_l}), \hspace{2em}V_n\in SU(2), 
  \label{gauge}
\ee
the same gauge transformation as in lattice Yang-Mils theory.  

\begin{figure}[t]
\centerline{\includegraphics[scale=0.3]{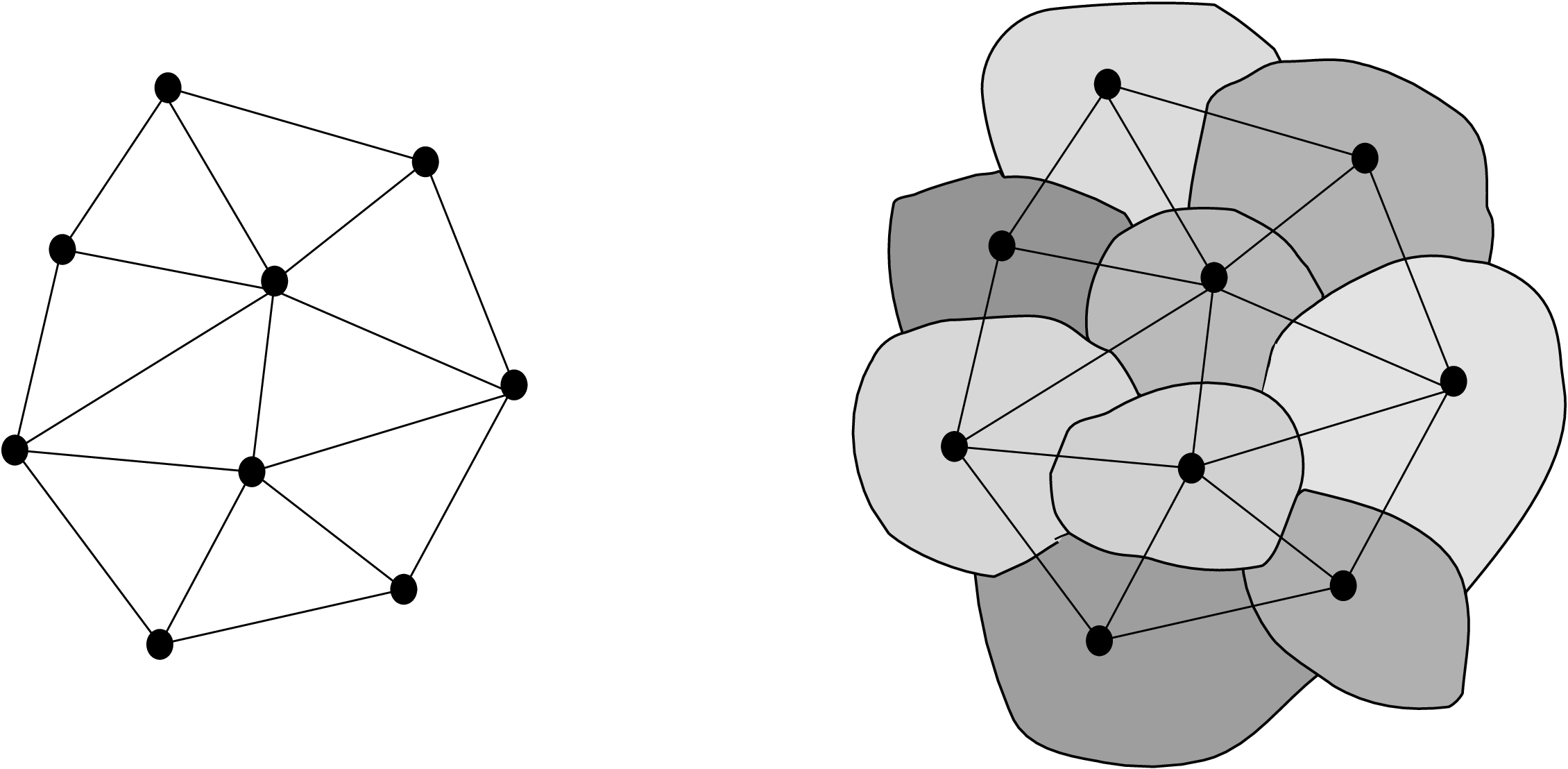}}
$\hspace{-15em}|\Gamma, j_l,v_n\rangle$
\label{uno}
\caption{{``Granular" space}. on the left, a graph. On the right, each node $n$ of the graph determines a ``quantum" or ``grain" of space.} 
\end{figure}

States in  $H_\Gamma$ represent quantum excitation of space formed by (at most) $N$ ``quanta of space". The notion of ``quantum of space" is basic in loop gravity. It indicates a quantum excitation of the gravitational field, in the same sense in which a photon is a quantum excitation of the electromagnetic field.  But there is an essential difference between the two cases, which reflects the difference between the electromagnetic field in Maxwell theory, and the gravitational field in general relativity: while the former  lives over a fixed (Minkowski) metric spacetime, the second represents itself spacetime.  Accordingly, a photon is a particle that ``lives in space", that is, it carries a momentum quantum number $\vec k$, or equivalently a positions quantum number $\vec x$, determining the spatial localization of the photon with respect to the underlying metric space.  The quanta of loop gravity, instead, are not localized in space. Rather, they define space themselves, as do classical solutions of the Einstein equations.  

More precisely, the $N$ ``quanta of space" are only localized with respect to one another: the links of the graph indicates ``who is next to whom", that is, the adjacency link that defines the spatial relations among the $N$ quanta.   (See the right panel in Figure 1.) Thus, these quanta carry no quantum number such as momentum $\vec k$ or position $\vec x$.\footnote{In the formal derivation of the quantum theory from general relativity, that I will sketch later on, the dependence on $\vec x$ is washed away by imposing the 3d diffeomorphism invariance of the quantum states.} 
Rather, they carry quantum numbers that define a quantized geometry, namely a quantized version of the information contained in a classical (three-dimensional) metric. 

The way this happens is elegant, and follows from a key theorem due to Roger Penrose, called the spin-geometry theorem, which is at the root of loop gravity \cite{Penrose2}.  I give here an extended version of this theorem,  based on a theorem proven by Minkowski in 1897 \cite{Minkowski:1897uq}.  

Consider the natural momentum operators on $L_2[SU(2)^L]$. These are given by the left invariant derivative operators $\vec L_l$ acting on the group elements $h_l\in SU(2)$, for each link $l$. The operator $\vec L_l$ is a 3d vector, since it lives in the $SU(2)$ algebra. By gauge invariance, when acting on gauge invariant states satisfying \eqref{gauge},
\be
   \sum_{l\in n} \vec L_l =0
   \label{gauss}
\ee
where the sum is over all links bounded by $n$ (taken to be outgoing). The operators $\vec L_l$ themselves are not gauge invariant, but gauge-invariant operators are easily defined by taking scalar products: the gauge invariant operator
\be
  G_{ll'} =\vec L_l\vec L_{l'}
\ee
where $l$ and $l'$ share the same source, is called the Penrose metric operator. 
Its diagonal elements   
\be
  A_l^2 =\vec L_l\vec L_{l}
\ee
are the Casimirs of the $SU(2)$ group associated to the link $l$ and play an important role in the theory.

Consider the $v\times v$ Penrose operators $G_{ll'}$ associated to a single node $n$ with $v$ adjacent links $l,l',l'',...$, namely to a single quantum of space. The theorem  states that the above equations are sufficient to guarantee the existence of a (flat) polyhedron with $v$ faces, such that the area of its faces is $A_l$ and $G_{ll'}=A_lA_{l'}\cos\theta_{ll'}$, where $\theta_{ll'}$ is the angle between the normals to the faces $l$ and $l'$. More precisely,  there exist a $3\times 3$ metric tensor $g_{ab}, a,b=1,2,3$ and normal to the faces $\vec n_l$, such that 
\be
G_{ll'} =g_{ab} n_l^an_{l'}^b
\ee
and the length of the these normals is equal to the area of the face. In other words, the algebraic structure of the momentum operators in $H_\Gamma$ determine the existence of a metric at each node and therefore equips each quantum of space with a geometry. 

Notice that, in general, gluing flat polyhedra does not yield a flat space. It may give a cellular discretization of a curved manifold, where curvature resides on the bones of the decomposition, as in Regge calculus.\footnote{To be sure, glued polyhedra define a discretization more singular than a Regge geometry, in general, because the later assumes continuity of the metric across faces. Such discretized metrics are called ``twisted geometries" \cite{Freidel:2010aq}.}  Thus, the momenta operators on $H_\Gamma$ describe a discretization of a curved space.

In particular, $A_l$ can be interpreted as the area of the surface bounding two quanta of space connected by the link $l$, and $\theta_{ll'}$ as the angle between the normals of two faces.\footnote{Such areas and angles can be chosen as variables for describing the gravitational field \cite{Dittrich:2008va}. The angle operator has first been explored in the loop context in \cite{Major:1999mc}.}  This is the core of Penrose's spin geometry theorem.  We can then immediately write also the volume of a quantum of space using standard geometrical relations. For instance, for a 4-valent node $n$, bounding the links $l_1,...,l_4$ the volume operator $V_{n}$ is given by the expression for the volume of a tetrahedron
\be 
V^2_{n}= |\vec L_{l_1}\cdot(\vec L_{l_2}\times \vec L_{l_3})|,
\ee
where gauge invariance \eqref{gauge} at the node ensures that this definition does not depend on which triple of links is chosen. 

The Penrose metric operator components do not commute with one another, therefore the geometry they define is never sharp, in the sense of quantum theory. It is always a genuinely ``quantum" geometry, where the angles between faces do not commute, as the components of the spin of a rotating quantum particle.  

Area and Volume $(A_l,V_n)$, however, form a complete set of  commuting operators in $H_\Gamma$, in the sense of Dirac. Therefore they define  orthonormal basis where these operators are diagonal.  The orthonormal basis that diagonalizes Area and Volume has been found and is called  the \emph{spin-network} basis.  This basis can be obtained via the Peter-Weyl theorem and it is given by 
\be
\psi_{\Gamma, j_l,v_n}(h_l)=\big\langle \otimes_l\,d_{j_l}\, D^{j_l}(h_l)\; \big |\; \otimes_n v_n\; \big\rangle {}_\Gamma
\label{sn}
\ee
where $D^{j_l}(h_l)$ is the Wigner matrix in the spin-$j$ representation, $v_n$ is an invariant tensor in the tensor product of the representations associated to each node
(an ``intertwiner") and $\langle \cdot |\cdot  \rangle_\Gamma$ indicates the pattern of index contraction between the indices of the matrix elements and those of the intertwiners given by the structure of the graph.\footnote{Both tensor products live in 
\be
H_\Gamma\subset L_2[(SU2)^L]=\bigoplus_{j_l} \bigotimes_l\, {\cal V}_{j_l}\otimes{\cal V}_{j_l}=\bigoplus_{j_l} 
\bigotimes_n\bigotimes_{e\in \partial n}{\cal V}_{j_l}.
\ee
where ${\cal V}_j$ is the $SU2$ spin-$j$ representation space, here identified with its dual.}  Since the Area is the $SU(2)$ Casimir, while the volume is diagonal in the intertwiner spaces, the spin $j_l$ is easily recognized as the Area quantum number and $v_n$ is the Volume quantum number.

A crucial result is that the area eigenvalues present a gap between zero and the lowest non-vanishing eigenvalue.  The area gap (in units that I discuss later on)  is the lowest non-vanishing eigenvalue of the $SU(2)$ Casimir
\be
     a_o=\sqrt{\frac12\left(\frac12+1\right)}=\frac{\sqrt 3}2
     \label{gap}
\ee
This ``area gap" is directly responsible for the ultraviolet finiteness of the theory, for the quantum cosmology phenomenology, and other things. 

Each Hilbert space $H_\Gamma$ already includes all the states in $H_{\Gamma'}$ where $\Gamma'$ is a subgraph of $\Gamma$: these are the states that do not depend on $h_l$ if the link $l$ is in $\Gamma$ but not in $\Gamma'$. Equivalently, states with $j_l=0$ on these links. Therefore $H_\Gamma$ for a graph $\Gamma$ with $N$ nodes, is the analog of the subspace $H_N$ of Fock space containing all states that have at most $N$ particles.  

Thus, an orthonormal basis of quantum states of space is labelled by three sets of ``quantum numbers": an abstract graph $\Gamma$; a half-integer coloring $j_l$ of the links of the graph with nontrivial irreducible unitary representations of $SU(2)$ for each link, and an intertwiner $v_n$ (eigenstate of the volume operator at the node) for each node. They represent quantum states of space, or quantum 3-geometries. $\Gamma$ gives the adjacency relations between the quanta of space, $v_n$ gives the volume of the quanta of space, $j_l$ gives the area of the faces that separate two quanta of space.  The states $|{\Gamma, j_l, v_n}\rangle$ labelled by these quantum numbers are called ``spin network states", and are the fundamental tools of the theory.  This rich geometrical structure is not imposed by hand: it is implicit in the algebra of the momenta operators on $H_\Gamma$.%
\footnote{Which in turn, as we shall see later on, derives directly from a canonical quantization of general relativity.}

The geometry represented by a state  $|{\Gamma, j_l, v_n}\rangle$ is a \emph{quantum} geometry for three distinct reasons. 
\begin{enumerate}
\item[i.] It is discrete. The relevant quantum discreteness is not the fact that the continuous geometry has been discretized ---this is just a truncation of the degrees of freedom of the theory. It is the fact that area and volume are quantized and their spectrum turns out to be discrete. It is the same for the electromagnetic field. The relevant quantum discreteness is not that there are discrete modes for the field in a box: it is that the energy of these modes is quantized.
\item[ii.] The components of the Penrose metric operator do not commute. Therefore the spin network basis diagonalizes only a subset of the geometrical observables, precisely like the $|j,m\rangle$ basis of a particle with spin. Angles between the faces of the polyhedra are quantum spread in this basis.  
\item[iii.] A generic state of the geometry is not a spin network state: it is a linear superposition of spin networks. In particular, the \emph{extrinsic} curvature of the 3-geometry\footnote{In canonical general relativity the extrinsic curvature of a spacelike surface is the quantity canonically conjugate to the intrinsic geometry of the surface.}, which, as we shall see later on, is captured by the group elements $h_l$, is completely quantum spread in the spin network basis. It is possible to construct coherent states in $H_\Gamma$ that are peaked on a given intrinsic as well as extrinsic geometry, and minimize the quantum spread of both.  A technology for defining these semiclassical states in $H_\Gamma$ has been developed by a number of authors, yielding beautiful mathematical developments that unfortunately I do not have space to cover here.  See for instance \cite{Thiemann:2002vj,Livine:2007mr,Freidel:2010bw,Bianchi:2009ky,Bianchi:2010gc}.
\end{enumerate}

States in $H_\Gamma$ are analog to $n$-particle states, since they include $N$ quanta of space.  But they are also extremely similar to the states of (hamiltonian) lattice gauge theory, living on a 3d lattice. Indeed, $H_\Gamma$ is nothing else than the standard canonical Hilbert space of an $SU(2)$ lattice Yang-Mills theory defined on the lattice $\Gamma$ \cite{KoguthSusskind}.

This convergence between the perturbative-QED picture and the lattice-QCD picture follows directly from the key physics of general relativity:  the fact that it describes the physics of space itself. Indeed, the lattice sites of lattice QCD are small regions of space; according to general relativity, these are excitations of the gravitational field, therefore they are themselves quanta of a (generally covariant) quantum field theory.  An $N$-quanta state of gravity has therefore the same structure as a Yang-Mills state on a lattice with $N$ sites. I find such a convergence between the perturbative-QED and the lattice-QCD pictures to be a beautiful feature of loop gravity. 

This is the quantum geometry at the basis of loop gravity. Let me now move to the transition amplitudes between quantum states of geometry. 

\subsection{Transition amplitudes}\label{dyn}

\begin{figure}[ht]
\centerline{\includegraphics[scale=0.2]{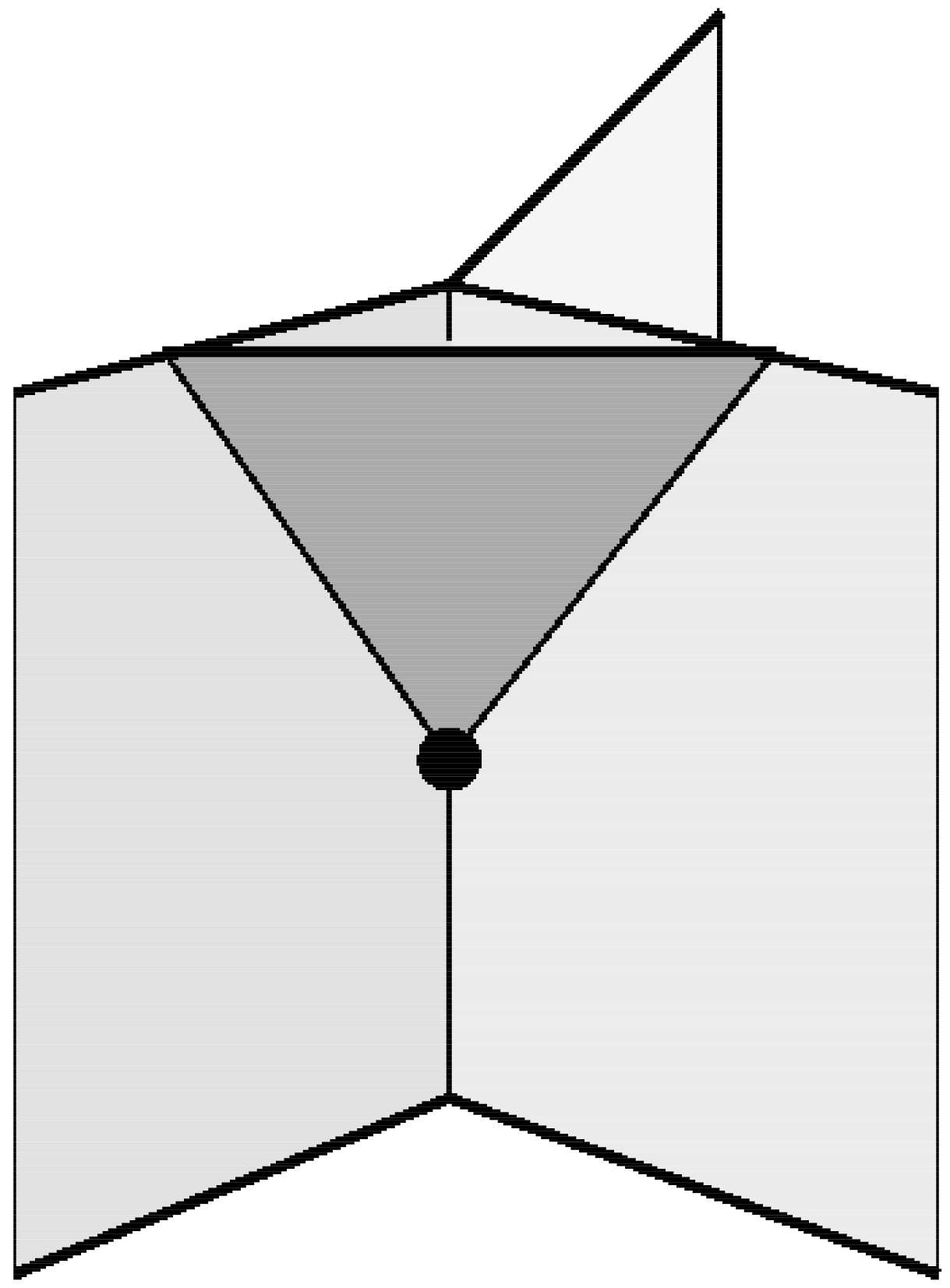}}
\caption{A two-complex with one internal vertex.}
\label{due}
\end{figure}

The transition amplitudes defined by equations (\ref{int1},\ref{va},\ref{K}) depend on a two-complex $\cal C$, or ``foam". The two-complex is the analog of a Feynman diagram. It is more complicated than a Feynman diagram, because states are not just quanta (particles), but are quanta (nodes) with their adjacency relations (links). The edges of the foam are the propagators of the nodes and the faces of the foam are the propagators of the links. See Figure \ref{due}.

 The dynamics is coded into the vertex amplitude $A_v$ given in \eqref{va}. This is the quantum-gravity analog of the QED vertex amplitude. 
\be
   \raisebox{-.5cm}{\includegraphics[scale=0.4]{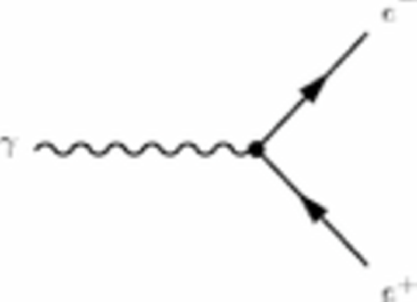}}
   =
   e\ \gamma_\mu^{AB}\ \delta (p_{1}\!\!+\!p_{2}\!\!+\! k )\nonumber
\ee
The simple QED vertex amplitude codes the nonlinear Maxwell-Dirac equation. In the same way, the vertex amplitude%
\footnote{Which is finite \cite{Engle:2008ev,Kaminski:2010qb}.}
 \eqref{va} codes the nonlinear Einstein equations. 
 
There are several elements of evidence for this claim. Consider a small 3-sphere surrounding the vertex, and let $\Gamma_v$ be the graph formed by the intersection between the 3-sphere and the foam. The labels on the foam define a quantum state in $H_{\Gamma_v}$, which has an interpretation as a discrete 3d geometry $q$, as discussed in the previous section. Consider a \emph{flat} spacetime region bounded by such a 3d geometry.  Recall that in Regge calculus, a curved spacetime is discretized in terms of a cellular decomposition, where each 4-cell is flat, and curvature is concentrated on the 2-cells. The action of general relativity is then discretized as a sum of terms $S_R(q)$, associated to each 4-cell, that depend on the discrete boundary geometry $q$ of the 4-cell.  Now, it  turns out \cite{Barrett:2009mw,Conrady:2009px,Bianchi:2009ky}, remarkably, that 
\be
A_v(q) \sim e^{iS_R(q)}
\ee
where the equality holds in the limit of large quantum numbers.\footnote{For the moment, this considerable result has been proven rigorously only in the case of 5-valent vertices.} It follows that the integral \eqref{int1} is a Regge-like discretization of the formal ``functional integral over 4-geometries"
\be
Z(q)=\int_{\partial g=q} \ dg\ e^{iS[g]}. 
\ee
where $S[g]$ is the Einstein Hilbert action.  That is, the equations  (\ref{int1},\ref{va},\ref{K}) are nothing else than a way of constructing the path integral for general relativity, but taking into account the discrete nature of quantum geometry, which is coded in the compactness of $SU(2)$.

Notice that here again the structure of loop gravity combines the formal structures of QED and lattice QCD.  On the one hand the transition amplitude \eqref{int1} for a fixed two-complex $\cal C$ is similar to a Feynman graph, where a finite number of particles (here quanta of space) is involved.  On the other, the same expression can be interpreted as a lattice discretization of the Feynman path integral definition of a quantum field theory, as the one used in lattice QCD, where the truncation is not on the number of particles, but rather on the finiteness of the discretization of spacetime.  The two pictures merge in quantum gravity because the individual ``particles" of background-independent quantum gravity are also ``spacetime" chunks.  Therefore to truncate in particle number is like truncating to a finite discretization of spacetime.

In Feynman diagrams, we increase the approximation by \emph{summing} over graphs.  In lattice QCD, we increase the approximation by \emph{refining} the lattice.  Which of the two procedures gives here the continuum limit? Remarkably, the two procedures are in fact equivalent \cite{Rovelli:2010qx}. By going to the spin network basis and including appropriate combinatorial factors (I do not enter here these more technical details of the theory), we can re-express the amplitude of a foam as a sum over amplitudes over all its subfoams where the sum over spins in \eqref{K} does not include $j=0$. In other words, all sub-complexes contributions are already taken into account by the amplitude on a given complex. Formally, the continuum limit can be defined by the limit 
\be
       Z(h_l)=\lim_{{\cal C}\to\infty}Z_{\cal C}(h_l)
       \label{cl}
\ee
which is well defines in the sense of nets\footnote{$\forall\epsilon,\exists{\cal C}'$ such that ${\cal C}>{\cal C}'$ implies $|Z(h_l)-Z_{\cal C}(h_l)|<\epsilon$.} because two-complexes form a partially ordered set with upper bound.\footnote{$\forall{\cal C},{\cal C}', \exists {\cal C}''$ such that ${\cal C}''>{\cal C}$ and ${\cal C}''>{\cal C}'$.}

In principle, the amplitudes \eqref{int1} can be used to compute all quantum gravity phenomena.  For instance, in a cosmological context, the transition amplitudes have been used to directly compute the quantum evolution of a ``universe", meaning the large scale modes of a compact geometry \cite{Bianchi:2010zs}.  A standard technique, which works also in the case of flat space quantum field theory, can be used to translate the field-to-field transition amplitude into the computation of $n$-point functions over a given field configuration \cite{Mattei:2005cm}.  The transition amplitudes  \eqref{int1} can be used to compute the $n$-point function of a graviton, in particular the graviton propagator, over a given background geometry \cite{Rovelli:2005yj,Modesto:2005sj,Bianchi:2006uf,Bianchi:2009ri}.  I give more details on some of these results later on, in chapter \ref{applications}.  

\subsection{Scale}

As defined so far, there is no scale in the theory. Everything has been defined in terms of dimensionless quantities.  In order to read areas and volumes in terms of physical units (centimeters), we have to transform the results above into centimeters.  

The theory lives at a certain scale. This is for instance the scale of the physical value of the area gap \eqref{gap}. Let me call $L_{loop}$ the centimeters value of the units in which \eqref{gap} holds.  $L_{loop}$ is the intrinsic scale of the theory, the scale at which spacetime becomes discrete. 

What is this value? It can be deduced by deriving the kinematics I have given here from a canonical quantization of general relativity (see below). Since the gravitational action is divided by (16$\pi$) the Newton constant $G$, the momentum conjugate to the gravitational connection is proportional to $G$. Promoting Poisson brackets to quantum operators leads to a momentum operator proportional to $\hbar G$.  This implies that the $\vec L_l$ metric operator, written in physical units, is $8\pi\hbar G\gamma$ times the dimensionless operator. That is 
\be
   L^2_{loop}=8\pi\hbar G\gamma.
\ee
In other words, the area gap reads  
\be
     a_o=8\pi\hbar G\gamma\  \frac{\sqrt 3}2
     \label{gap2}
\ee
in physical units. 

This argument, however, must perhaps be taken with caution, not only because of the uncertainty in the value of $\gamma$, but especially because of possible (likely) scaling effects of the Newton constant between its Planck scale value $G_{Planck}$ and its large-distance value $G$. On the other hand, we know from perturbation theory that this scaling is probably not large (unless there exists a great number of particles), see \cite{Larsen:1995ax,Calmet:2010qq} and references therein, and I'd say that we can  assume that the order of magnitude of $L_{loop}$ is still $L_{loop}\sim L_{Planck}:=\hbar G$. This is the scale of loop gravity. 

It is important to stress the substantial difference in which scale appears in quantum gravity and in lattice QCD. Lattice QCD is defined at a given scale $a$, which is the distance between lattice sites in the background metric. This scale affects (indirectly) the action. The continuum limit is then taken by a double limit: $a$ is taken to zero (more precisely, the coupling constant in the action is taken to its critical value) and  at the same time the lattice is refined (meaning: the number of lattice sites is increased.) In the limit, correlation functions extend over an infinite number of lattice sites, while the physical distance between lattice sites is taken to zero. 

In quantum gravity, instead, there is no lattice size $a$. The graph $\Gamma$ on which a truncation of the theory is defined carries no length dimension. This is the same as for the coordinates in general relativity, which have no metric significance by themselves.  Distances are determined by the field variables, namely by the quantum states living on the graph. In addition, these distances can never become smaller than the Planck scale. It follows that the continuum limit is intrinsically different, and in fact much simpler, than in lattice QCD. There is no lattice spacing to take to zero, but only the refinement of the two-complex to take to infinity. Therefore the continuum limit is entirely captured by \eqref{cl}.\footnote{This does not change the fact that the effect of radiative corrections can be subsumed under a dressing of the vertex amplitude, improving the regime where the expansion  \eqref{cl} is effective.}

\vskip.5cm\centerline{---}\vskip.5cm

This concludes the definition of the theory. I have given here this definition without mentioning how it can be ``derived" from a quantization of classical general relativity. This is like defining QED by giving its Hilbert space of free electrons and photons and its Feynman rules, mentioning neither canonical nor path integral quantization. A reason for this choice is that I wanted to present the theory compactly. A second reason is that one of the main beauties of the theory is that it can be derived with rather different techniques, that involve different kinds of math and different physical ideas. The credibility of the end result is reinforced by the convergence, but each derivation ``misses" some aspects of the whole.  In chapter \ref{history} below I will briefly summarize the main ones of these derivations. Before this, however, let me discuss what is this theory meant to be good for.

\section{The problem addressed}\label{problem}

What is the problem that the theory described above means to solve?  Let me begin by giving a list of problems that the theory does \emph{not} intend to solve, because thinking about these problems has often raised confusion in discussing the merits of the theory.
\begin{description}
\item[\it{Unification}.] The idea is often put forward that a quantum theory of gravity must be also a theory that unifies all forces. There are hints that this might be the case. For instance the fact that the running of the constants of the Standard Model appears to converge not too far from the Planck scale, or that fermions and boson divergences tend to cancel in supersymmetric theories.  Often in the history of physics two different problems have been solved at once (combing electricity and magnetism and understanding the nature of light) and the temptation to try this again is reasonable. 
But, even more often, proclaimed hopes to solve two problems at once have turned out to be false hopes (finding a theory for the strong interactions and at the same time getting rid of renormalization theory, for example. This was an unquestioned mantra when I was a student, and it turned out to be a false piste.) Loop quantum gravity is resolutely a theory of quantum gravity that does not address the unification problem. In this, it is like QED, or, more precisely, QCD: a quantum field theory for a certain interaction, which can be coupled to other interactions, but is also consistent by itself.  The philosophy underlying loop gravity is that we are not near the end of physics, we better not dream of a final theory of everything, and we better solve one problem at the time, which is hard enough.  

\item[\it{Meaning of quantum mechanics}.] \hfil Hopes have been voiced that a quantum theory of gravity will clarify the mysteries of quantum theory.  This is not the case of loop gravity, which uses standard quantum theory (in whatever interpretation is your favorite one), only slightly generalized to make room for the peculiar way temporal evolution is described in general relativity. 

\item[\it{Quantum cosmology as of ``theory of all things"}.] \hfil Loop gravity definitely addresses quantum cosmology (see below), but only in the sense of ``the theory of the large scale degrees of freedom of the universe" (the scale factor, the large scales fluctuations, and so on.)  These degrees of freedom form a \emph{subset} of all degrees of freedom (quantum cosmology does not describe this computer on which I am writing). Thus, loop gravity has nothing to contribute to problems such as ``doing the quantum physics without observer" or similar. 
\end{description}

So, what is the problem that loop gravity does address? Simply: Is there a consistent quantum field theory whose classical limit is general relativity?

If we want to compute the gravitational scattering amplitude of two point-particles, as a function of their center of mass energy $E$ and impact parameter $b$, we can do so using (non-renormalizable)  perturbative quantum general relativity used as an effective theory.  However, the predictivity of the theory breaks down when $E$ increases and becomes of the order of $b$ (in Planck units).  So, current established theory is incapable of predicting what happens to particles that scatter at that energy.  The aim of loop gravity is to fill up this lack of predictivity.\footnote{It is interesting to observe that if $E$ is much larger than $b$ the established theories again allow us to predict the scattering amplitude. Essentially what happens is that a black hole is formed and the event can be described using classical general relativity. Therefore the regime properly covered by quantum gravity is a narrow wedge around $E\sim b$.} The lack of predictivity of current established theories is particularly relevant in certain well known physical situations, the main ones of which are:
\begin{description}
\item[\em Early cosmology,] where the so called ``Big Bang" is not an event predicted by current theories, but just a sign that current theories become insufficient because quantum properties of spacetime dominate. 
\item[\em Black hole thermodynamics:] the tantalizing semiclassical results leading to the Bekenstein-Hawking entropy call for a first-principle derivation. 
\item[\em Short-scale structure of physical space.] It is reasonable to expect the description of spacetime as a pseudo-Riemannian manifold to break down, for the same reason for which the description of the electromagnetic field as a smooth field breaks down at the scale of the individual photons.
\end{description}

There are then three major theoretical and conceptual problems that theory addresses:
\begin{description}
\item[\em Quantum geometry:] What is a physical ``quantum space"? That is, what is the mathematics that describes the quantum spacetime metric? 
\item[\em Ultraviolet divergences of quantum field theory.] This is a major open problem in non-gravitational contexts. But it is a problem physically related to quantum gravity because the ultraviolet divergences appear in the calculations as effects of ultra-short trans-Planckian modes of the field. If physical space has a quantum discreteness at small scale, these divergences should disappear. 
\item[\em General covariant quantum field theory.]\hfil
Loop gravity ``takes seriously"  general relativity, and explores the possibility that the symmetry on which general relativity is based (general covariance) holds beyond the classical domain. Since standard quantum field theory is defined on a metric manifold, this means that the problem is to find a radical generalization of quantum field theory, consistent with full general covariance and with the physical absence of a background metric structure. In other words, loop gravity, before being a quantum theory of general relativity, is the attempt to define a {\em general covariant quantum field theory}.
\end{description}

There are standard arguments against the possibility of finding a consistent quantum theory of general relativity alone.  These arguments have motivated the exploration of supersymmetry and other extensions of the theory. They are also often mentioned to claim  that quantum general relativity cannot be consistent unless suitably coupled to other fields. These arguments, however, hold only in the context of standard \emph{local} field theory, where fields operators are defined on a spacetime metric  manifold. They are all circumvented by moving up to the context of a general-covariant quantum field theory.

All these problems \emph{are} addressed by loop gravity, and definite answers are given. These will be discussed in the section  \ref{assessment}.  First, however, I give a (incomplete) historical account of the development of the theory. My main purpose is to explain how the theory defined in chapter \ref{art} can be (and has been) ``derived" from classical general relativity. 

\section{The roads to LQG}\label{history}

\subsection{Ante loops}

Loop gravity is grounded on a number of ideas that constitute the background thinking of many quantum gravity researches:
\begin{description}[\setlabelstyle{\itshape}]
\item[Sum over geometries.] The idea that quantum gravity could be defined by an expression like
\be
Z=\int dg\ e^{iS[g]}. 
\label{Z}
\ee
first appeared in the fifties in a paper by Charles Misner, where Misner attributes it to John Wheeler \cite{Misner:1957fk}. The idea was later extensively developed (in its Euclidean version) by Stephen Hawking and his collaborators.  The problem with \eqref{Z} is of course that it is very badly defined and attempts to define it perturbatively lead to the standard ultraviolet divergences. In loop quantum gravity, \eqref{Z} is concretely realized by equation \eqref{int1}, which can be interpreted as providing a family of approximations to \eqref{Z}. The key difference between \eqref{Z} and  \eqref{int1} is not the discretization: other discretizations of  \eqref{Z} suffer from the same ultraviolet difficulties as  \eqref{Z}. The key difference comes from the fact that the integrals overs $SU(2)$ in   \eqref{int1} are over a \emph{compact} space. Physically, this means that the conjugate variable is discrete. The conjugate variables to the $SU(2)$ group elements are the ones that describe the three geometry, illustrated above in section \ref{spin}, and the geometry eigenvalues are \emph{discrete}. Thus, \eqref{int1} can be viewed as like a more precise formulation of \eqref{Z}, obtained by inserting resolutions of the identity in the evolution operator, \`a la Feynman, where the resolution of the identity is a sum over intermediate states of the 3-geometry, and these intermediate states are labeled by \emph{discrete} quantum numbers. In particular, there is a gap between the minimal area and zero. Which means that ultra-short, trans-Planckian degrees of freedom never appear in  \eqref{int1}. This is the physical reason for the ultraviolet finiteness of loop gravity: the path integral is over quantum configurations of spacetime and takes into account that in these trans-Planckian degrees of freedom do not appear ---for the same reason for which energies smaller than $E_o=\frac12 \hbar\omega$ do not appear in the physics of the harmonic oscillator. 

\item[Penrose spin geometry theorem,] \cite{Penrose2}, which I have already discussed above. A posteriori, it is astonishing that Penrose guessed a structure for quantum space, which is the same as that emerged from a follow-the-rules quantisation of general relativity (see below).

\item[Regge calculus] \cite{Regge:1961px}.  The idea of Regge is that general relativity admits a natural truncation in terms of Regge metrics. A Regge metric is a metric which is flat everywhere except on $(n-2)$-dimensional defects (triangles in 4d and bones in 3d), where curvature is concentrated and constant. The geometry of a Regge manifold is therefore captured by the flat geometry of a manifold with $(n-2)$-d defects plus the holonomy of the Levi-Civita connection (or the spin connection) around the defects. The graphs and the two-complexes of loop gravity are essentially a description of a triangulated Regge manifold. In 3d (space), the graph is the 1-skeleton of the dual to the 3d spatial manifold with defects: loops in the graph wrap around the Regge bones.  In 4d (spacetime) the two-complex is the two-skeleton of the dual of a 4d cellular decomposition. Again, the faces of the two complex wrap around the Regge triangles where curvature is concentrated.  Thus, loop gravity is like a ``topological" field theory where curvature is everywhere flat, except that it is defined on a manifold with defects, and therefore there is room for curvature, \`a la Regge, on these defects \cite{Immirzi:1995yk,Immirzi:1996dr,Bianchi:2009tj}. This is also the picture that relates quantum gravity and Atiyah's topological field theories, as I will discuss later on.  Concretely, as I have mentioned above, the transition amplitudes \eqref{int1} approximate a sum over Regge geometries of the exponential of the Regge action.  However, this is not quantum Regge calculus, because the quantization of geometry at small scale cuts the trans-Planckian degrees of freedom, which are still there in quantum Regge calculus. 

\item[Matrix models] were studied in the eighties as a way to define quantum gravity in two dimensions \cite{David:1985vn,Gross:1989vs,Ambjorn:1985az,Kazakov:1985ea,Brezin:1977sv}. They were generalized to 3d by Boulatov  \cite{Boulatov:1992vp} and to 4d by Ooguri \cite{Ooguri:1992eb}. The Ooguri theory is the model over which the spinfoam dynamics has later been constructed. I illustrate this later on. 

\item[Wheeler-deWitt equation.] A formal canonical quantization of general relativity leads to the Wheeler-deWitt equation, or the ``Schr\"odinger equation of general relativity", a (system of infinite coupled) functional equation(s) for a `wave function" $\psi(q)$ of the three-geometry $q$, or a function of the three dimensional metric invariant under 3d diffeomorphisms.    A study of the Wheeler deWitt equation is the path that has first led to loop quantum gravity. In fact, the early interest for the theory raised sharply at the announcement that it was possible to write quantum states that are invariant under 3d diff and solve the Wheeler-deWitt equation \cite{Rovelli:1989za}. These states are the ``loop states" from which loop gravity was born. The Wheeler-deWitt equation can be formulated with precision in the canonical version of loop gravity \cite{ThiemannBook}. 

\item[Ashtekar variables.] The main ingredient that allowed the birth of loop gravity is the discovery by Abhay Ashtekar of a beautiful and clean reformulation of general relativity in terms of an $SU(2)$ connection variable $A$ \cite{Ashtekar86}. 
 The Ashtekar variables, both in their original complex version and in later real versions remain the main tool of the theory.
 
 The shift from metric to connection variables introduced by Ashtekar opened the possibility to incorporate techniques from YM theory into quantum gravity, with immediate results, as I illustrate below.
\end{description}

\subsection{From the connection to loops}

In 1987, Ted Jacobson and Lee Smolin published a surprising result \cite{JacobsonSmolin}: by writing the Wheeler-deWitt equation in terms of Ashtekar variables, it was possible to find exact solutions to the equation. These were simply given by Wilson loops --traces of path ordered line integrals of the Ashtekar connection--, along closed lines provided that the lines did not self-intersect. (Below I give a simple-minded derivation of this result.)

The Jacobson-Smolin  solutions are not physical states of quantum gravity, because they fail to satisfy the second equation of canonical quantum gravity, which demands states to be invariant under 3d diffeomorphisms. Immediately, Smolin started to speculate that since loops up to diffeos mean knots, knots could play a role in quantum gravity.  

The idea that started the loop representation of quantum gravity was then the following. Let $A$ be the  Ashtekar connection and $\gamma$ a differentiable closed loop in space.%
\footnote{Beware: the symbol $\gamma$ is used in loop gravity with two unrelated meanings: (i) the Barbero-Immirzi parameter, (ii) a generic loop in space.}
  Let 
\be
\psi_\gamma[A]= Tr [Pe^{\int_\gamma A}]
\ee
where $Pe^{\int_\gamma A}$ is the parallel transport operator  of $A$ along $\gamma$ (a bit inappropriately called ``holonomy" in the jargon of the community). Then write 
\be
     \psi(\gamma)=\int dA \ Tr [Pe^{\int_\gamma A}] \ \psi(A)
\label{transform}
\ee  
where $dA$ is a diff-invariant measure on the space of connections. (A measure with this property was not known at the time, but was later constructed by Ashtekar and Lewandowski.)  The ``loop transform" \eqref{transform} can be viewed as a generalization of a Fourier transform that maps the representation of quantum states in $x$-space to the representation of quantum states in momentum space \cite{Rovelli:1991kx}.  Here, the transform maps the $\psi(A)$ representation of quantum states in $A$ space to the representation $\psi(\gamma)$ of quantum states in $\gamma$ space, namely in loop space. This change of basis is the one that has evolved into the modern change of basis \eqref{sn}. 

What are the advantages of going to the loop basis? Several. The first is that the diffeomorphism constraint can be immediately solved: it is solved by any $\psi(\gamma)$ that depends only on the knot-class of the loop $\gamma$. In other words, there is one independent solution for each knot. (Compare to the fact that the graphs $\Gamma$ used in section \ref{spin} are not imbedded in a space, but just combinatorial objects.) Second, all such states where the loop does not self-intersect are exact solutions of all equations of quantum gravity. This is the result that sparkled the early interest in the ``loop representation of quantum general relativity" \cite{Rovelli:1987df,Rovelli:1989za}. 

So many solution to the quantum dynamics of general relativity sounded a bit too easy. In fact, it soon turned out that the ``good" solutions where still missing. The hint was already in the Jacobson-Smolin paper, where it was shown that solutions can be obtained if the loops do not intersect, but also for certain appropriate combinations of intersecting loops. Thus, there are other solutions beside the ones defined by non intersecting loops: solutions where intersections between loops, or ``nodes", play a role.  Jorge Pullin was the first to realize that in fact all solutions without nodes correspond to 3-geometries with zero volume \cite{Brugmann:1991fk,Brugmann:1992uq}. (Later Renate Loll realized that the node must be at lest 4-valent to have volume \cite{Loll:1995wt}.) Thus the loops without nodes are solutions, but they describe a quantum space without volume.   It thus became rapidly clear that the nodes, namely intersections between loops, are not only unavoidable, but in fact essential to describe physical quantum geometry.  Nodes are where ``quanta of volume" sit.  These intersections later became the nodes $n$ of the graph $\Gamma$ of section \ref{spin}, above. A set of intersecting loops gives in fact an embedded graph, and the equivalence class of such embedded graphs under diffeomorphisms are labelled by abstract combinatorial graphs $\Gamma$, plus information about their knotting and linking properties.  

With the better comprehension of the theory, the knotting and linking properties have since moved to the background: the abstract graph structure is sufficient to describe quantum geometry.  To the point that in recent versions of the theory, such as the one I have given above, they are entirely dropped.  A line of research trying to see if they could nevertheless play a physical role in quantum gravity, and maybe even be used to code matter degrees of freedom, is nevertheless very active see \cite{BilsonThompson:2006yc,BilsonThompson:2009fh,Gurau:2010nd}.

The fact that the action of the Wheeler-deWitt operator $C$ vanishes away from the nodes retains its full importance also after the physical  significance of the nodes is understood.  The nontrivial action of  $C$ is confined to nodes. That is, the loop representation largely trivialize the quantum dynamics, reducing it to a combinatorial and discrete action on the nodes themselves. All modern versions of the quantum loop dynamics are still based on this fact. 

It is instructive to show a simple calculation to illustrate the fact that non-intersecting loops solve the Wheeler-de Witt equation and therefore the dynamics trivializes outside the nodes.  The hamiltonian constraint in (the euclidean version of) Ashtekar's theory is $C=F^{ij}_{ab}[A]E_i^aE_j^b$, where $F^{ij}_{ab}[A]$ is the curvature of the Ashtekar connection and $E_i^a$ is the variable conjugate to the Ashtekar connection, which has the geometrical interpretation of an (inverse densitized) triad field. Therefore the (euclidean) Wheeler-deWitt operator in the Ashtekar variables reads
\be
   C(x)=F^{ij}_{ab}[A](x)\frac{\delta}{\delta A^i_a(x)}\frac{\delta}{\delta A^j_b(x)}.
\label{C}
\ee
Let $\tau_i$ be a basis in the algebra of $SU(2)$ and $U_\gamma(A)=Pe^{\int_\gamma A}$. Using the identity
\begin{eqnarray}
\frac{\delta}{\delta A^i_a(x)}U_\gamma(A)&=&
\frac{\delta}{\delta A^i_a(x)}Pe^{\int ds \frac{d\gamma^a(s)}{ds} A^i_a(\gamma(s))\tau_i}
\\ \nonumber &=&\int ds \frac{d\gamma^a}{ds}\delta(x,\gamma(s))\;
U_{\gamma_1}(A)
\tau_i
U_{\gamma_2}(A)
\end{eqnarray}
where $\gamma_1$ and $\gamma_2$ are the two loops into which $\gamma$ is split by the point $x$, we can compute the action of $C$ on a loop state $\psi_\gamma$. This gives 
\begin{eqnarray} \label{naive}
  C(x)\psi_\gamma(A)&=&F^{ij}_{ab}[A](x)\frac{\delta}{\delta A^i_a(x)}\frac{\delta}{\delta A^j_b(x)}Tr [U_{\gamma}(A)] \nonumber \\
&  \sim& F^{ij}_{ab}[A](x)\int ds \frac{d\gamma^a}{ds}\delta(x,\gamma(s))
\\ \nonumber
&& \hspace*{-2mm} \int dt \frac{d\gamma^b}{dt} 
 \delta(x,\gamma(t))\  Tr [U_{\gamma_1}(A)
\tau_i \tau_j
U_{\gamma_2}(A)]
\end{eqnarray}
If $x$ is not a point where the loop self-intersects, $s=t$ because of the two delta functions, and the two tangents are parallel, but $F_{ab}$ is antisymmetric and this expression vanishes.  

By the way, this calculations also shows why nodes are needed to get volume. The volume density is the determinant of the triad, hence depends on 
\be
V\sim \epsilon_{abc} E^aE^bE^v\sim \epsilon_{abc}\frac{d\gamma^a}{ds} \frac{d\gamma^b}{dt} \frac{d\gamma^c}{dy} 
\ee
For this not to vanish there must be at least three distinct tangents of the loop in a single point of space (in fact, four, because if they are only three, \eqref{gauss} forces them to be linearly dependent and $V$ vanishes \cite{Loll:1995wt}). 

With this understood, the basic structure of the theory had been found. 
 But much remained to be done. The main open problems were understanding the action of $C$ on the nodes and making physical and mathematical sense of the funny ``loopy" geometry that seemed to emerge from the theory (What does it imply physically? Can any gauge theory be written in this representation?...).   But the early representation in terms of functionals of loops $\psi(\gamma)$ was cumbersome, and progress was slow.  The reason is that loop states are not independent, since holonomies of the connections are not independent.  Speed pic ked up as soon as this  technical problem was solved, but it took several years and numerous dead ends, before finding the good solutions. The solution, as is often the case, turned out to be very simple, and in fact more or less already known in the context of lattice gauge theory. 

\subsection{Spin networks}

The  solution is given by the spin network states.  Spin networks were introduced in the mid nineties, as a tool for solving the spectral problem for the operators representing the area of surfaces and volume of 3d regions, in the loop representation \cite{Rovelli:1995ac}, following a key hint by Jurek Lewandowski, who pointed out that the problem factorized on graphs. Essentially, the spin networks are given by the basis \eqref{sn} in the space state $H_\Gamma$ associated to a single graph $\Gamma$. These states are independent, form an orthonormal basis, and fully resolve the over-completeness of the old loop basis. 

With the step to graphs and spin networks, the original ``loops" are still there (a spin network state can be written as a linear combination of loop states), but have faded somewhat in the background.  Therefore ``Loop Quantum Gravity" is today somewhat of misnomer: it remains in spite of the fact that the theory is about graphs as much as loops ---like ``string theory" keeps its name in spite of the theory being today about branes as much as strings.

With the new tool provided by the spin networks, the most spectacular surprise provided by the theory was immediately found: the area of surfaces in space and the volume of regions in space have discrete spectra \cite{Rovelli:1994ge}. This requires perhaps a bit of physical explanation.  Area and volume are functions of the metric. The core physical discovery at the basis of classical general relativity is that the metric of space is nothing else than the gravitational field.  In a quantum theory of gravity, the gravitational field becomes an operator on a Hilbert space, like all other physical quantities in a quantum theory.    Then area and volume can be precisely constructed from the expression for area and volume as functions of the metric, namely the gravitational field.  Their spectra can be computed \cite{Rovelli:1994ge,Loll:1995wt,Ashtekar:1996eg,Ashtekar:1997fb}, and turn out to be discrete. For instance, the spectrum of the area is 
\be 
          A=8\pi\gamma\hbar G\sum_n\sqrt{j_n(j_n+1)}\label{spectrum}
\ee
for any finite set of half-integers $j_n$. 

This is indeed summarized in the construction given in section \ref{spin}, above, and is, in my opinion, the most solid result in loop gravity.  It is the basis for the ultraviolet finiteness of the theory. 

The first hint of such Planck-scale discreteness of space had in fact appeared much  before the solution of the spectral problem for Area and Volume. It came in a paper to which I am particularly attached \cite{Ashtekar:1992tm} where the first attempt to write ``semiclassical" states built from loop states, and capable of approximating a smooth geometry, was made.  In a completely unexpected way, it turned out that a state of this kind (a ``weave" state) could be constructed, but it would approximate a smooth geometry only down to the Planck scale, and not beyond it.  Trying to ``squeeze-in" more loops to better approximate smoothness failed: the mathematics refused to collaborate.  Contrary to expectations, to squeeze-in more loops turned out just to increase the physical size of space, instead of better approximate smoothness. In other words, individual loops appeared to behave as if they had an intrinsic Planck-scale minimal size.  This is precisely what later become clear with the geometrical spectra.

\subsection{Quantum geometry}

During the 90's, the structure described above has been posed on a solid mathematical basis by a remarkable series of constructions, crowned by two beautiful results.  The first is the definition of the Ashtekar-Lewandowski measure $d\mu_{AL}$. This is a measure on a functional space $\cal A$ of (suitably distributional)  connections $A$ on a manifold $M$. The measure is non-trivial, gauge-invariant, and invariant under 3d diffeomorphisms \cite{AshtekarLewandowski}.  This measure puts the transform \eqref{transform} on a solid footing and realizes the formal limit of of $H_\Gamma$ for arbitrary large $\Gamma$ as an $L_2$ space. More precisely, the formal limit
 \be
\lim_{\Gamma\to\infty} H_\Gamma 
 \ee
can be understood as living in the quotient of the space $L_2[{\cal A},d\mu_{AL}]$ by the diffeomorphism group.  $L_2[{\cal A},d\mu_{AL}]$  is then the natural mathematical home for the theory, and in particular for the geometric operators illustrated in section \ref{spin}.  Square integrable states $\psi[A]$ are typically ``cylindrical" functions of the form
\be
       \psi[A]=\psi_{e(\Gamma), j_l,v_n}[A]:=\psi_{\Gamma, j_l,v_n}(Pe^{\int_{e(l)} A})
\label{psiA}
\ee
where $\psi_{\Gamma, j_l,v_n}$ is the spin network function on $SU(2)^L$ defined in \eqref{sn} and $e:\Gamma\to M$ is a fixed embedding of $\Gamma$ into $M$. Thus, the characteristic of the Ashtekar-Lewandowski measure is precisely to be concentrated on ``polymer" geometries, in the sense that in \eqref{psiA} the field $A$ is smeared in just one dimension, and not in three dimensions, as square integrable states in functional representations of the Fock space typically are. Therefore the mathematical structure of loop gravity is intrinsically different from that of standard local quantum field theory. 

The second beautiful result is a uniqueness theorem, called the LOST theorem (from the initials of the authors that have proven one of its versions)  \cite{Fleischhack:2007mj,Lewandowski:2005jk}. The theorem states that under some very general conditions, the space $L_2[{\cal A},d\mu_{AL}]$ and the geometrical operators defined in section \ref{spin} are the only quantum representation of the corresponding Poisson algebra of classical general relativity quantities, up to unitary equivalence.  This extremely strong result, which recalls the Von Neumann uniqueness theorem of non-relativistic quantum theory, puts the kinematics of loop quantum gravity (Hilbert space and operators) on a very solid ground.  

The key hypothesis of the theorem is 3d diffeomorphism invariance. This fact shows where precisely quantum general relativity diverges from conventional quantum field theory, from the point of view of mathematical physics. It is diffeomorphism invariance that forces states to have the ``polymer" structure required by the Ashtekar-Lewandowski measure, at variance with quantum field theory on metric manifolds.  The difference can be physically understood as follows. A quantum field excitation on a metric space cannot be concentrated on a zero-volume set. Accordingly, a  normalizable Fock state must be smeared in 3d.   In a diffeomorphism-invariant theory, instead, the formal smearing over the manifold is already performed by the diffeomorphism gauge: gauge-invariant states are independent from their localization on the manifold.  The physical metric ``size" of the quanta, instead, is given by the state itself: the quanta of geometry are not point-like, since they have volume.  Intuitively speaking, each polymeric field excitation has already a Planck-scale ``thickness". This is the core of the mathematical representation of quantum geometry that underlines loop gravity. 

These developments have fully clarified the initial confusion on the nature of the loop representation. Questions like ``is there a physically-meaningful loop representation of QED?" have been answered. The answer is no; but there should be a physically-meaningful loop representation of the diff-invariant Maxwell-Dirac-Einstein system. And in fact, there is one \cite{Bianchi:2010ys}.

\subsection{Hamiltonian}

The Wheeler-deWitt operator $C$ annihilates states when there are no nodes.  How does it act on nodes? The full and rigorous construction of the operator $C$ is a complex story to which numerous people contributed. I only mention the main steps of this story. The original form of $C$ used to show that $C\psi_\gamma=0$ if $\gamma$ has no nodes, diverges on nodes. This is evident in the naive calculation \eqref{naive} where the two 3d delta functions can only be integrated against the five integrals $ds$, $dt$ and $d^3x$, leaving one delta function unsaturated. The proper definition of $C$ requires a regularization. Several regularizations were studied.

A key step was to realize that the limit in which the regularization cut-off is removed turned out to be finite under two conditions. First, that the limit is taken on a diffeomorphism invariant state. Second that the regularized operator is a version of $C$ with the proper density character \cite{Rovelli:1993bm}.  For instance, \eqref{C} does not have the proper density factor, which is 1, because it is a double density in $x$. One possibility, early explored, was to use the square root of $C$, but the square root has its own difficulty. Another was simply to divide $C$ by $\sqrt{\det E}$, to get back to a proper density. But the inverse of $\det E$ is a cumbersome operator. 
Then Thomas Thiemann proposed an ingenious solution, today known as ``Thiemann's trick": since $E^a$ and $A_a$ are conjugate variables, it follows that 
\be
   \tilde C:= \frac{E^aE^bF_{ab}[A]}{\sqrt{\det E}}=\epsilon^{abc}\{\!\int \!d^3\!x\sqrt{\det E},A_a \}F_{ab}[A]
\ee
where $\{\ , \}$ indicates Poisson brackets. This expression can be carried over to the quantum theory by replacing Poisson brackets with commutators and $\int d^3x\sqrt{\det E}$ with the volume operator. Therefore a density-1 Wheeler-deWitt operator $\tilde C$ could be defined in a relatively simple fashion.  This idea led to the detailed construction of Thiemann's Wheeler-deWitt operator, which defines the dynamics in the hamiltonian formulation of loop gravity \cite{Thiemann:1996aw,ThiemannBook}. 

The main merit of the hamiltonian version of the theory, in my opinion, is a remarkable proof of existence: a canonical quantization of general relativity, anomaly-free, general covariant, and without trans-planckian degrees of freedom, hence necessarily finite in the ultraviolet, can be defined.  A second major merit is that it has provided the ground for loop cosmology, which I discuss later on. 

A source of complication is the structure of the gauge invariance of the theory in the hamiltonian framework, which leads to the so called ``problem of time". The ``problem of time" is not anymore a conceptual problem in quantum gravity since the conceptual issues have been clarified, but remains a source of technical difficulties.  The problems can in principle be solved using the relational formalism. That is, defining observables not with respect to unphysical space time points but in terms of relations between dynamical fields. These ideas date back to the work of Bergmann and Komar \cite{Bergmann:1961zz,Bergmann:1961wa,Bergmann:1960wb} in the 1960s and were clarified and developed in the 1990s \cite{Rovelli:1990ph,Rovelli:2001bz,Dittrich:2004cb,Dittrich:2005kc}. But if the solution is straightforward in principle, and can be readily applied in simple contexts such as cosmology, it nevertheless becomes hard, as a practical matter, to implement in the full theory.  It is like having the Schr\"odinger equation of a protein, and not knowing what to do with it, because it is far too complicated.

One possibility of constructing relational observables is to couple the theory to effective matter fields and use these as reference systems, in order to formally circumvent the difficulties deriving from general covariance \cite{Rovelli:1990pi}.  
For instance, \cite{Dittrich:2006ee,Dittrich:2007jx} discuss a perturbative 
scheme to compute these observables and its application to 
perturbation theory around Minkowski space and cosmological perturbation 
theory for different choices of matter clocks. In principle, constructing observables is difficult, but in  \cite{Thiemann:2004wk,Thiemann:2006up,Giesel:2007wi,Giesel:2007wk,Giesel:2007wn}, Thiemann has observed that what is needed is just the 
algebra of those Dirac observables and the physical Hamiltonian which drives their 
physical time evolution, and this opens the way to actual calculations. See \cite{Giesel:2009jp,Giesel:2009at}.

A particularly intriguing development in this direction is the recent one in \cite{Domagala:2010bm}, where the idea \cite{Rovelli:1993bm} of keeping full 3d-diff invariance manifest but solving time reparametrization invariance by introducing a single scalar clock field has been implemented, with intriguing results.  The question is open whether these developments, which have provided a formal definition of the theory can also lead to a technique to extract from the theory effective physical information.

A second potential difficulty with the hamiltonian approach is the fact that the detailed construction of the Wheeler-deWitt  operator is intricate and a bit baroque, making questions of uniqueness particularly hard to address. The perception of it as more in the rigorous mathematical style of constructive field theory than in the direct computationally friendly language of theoretical physics may have contributed to growing involvement of a substantial part of the loop community with an alternative method of constructing the theory's dynamics.  It is to this alternative, seen as more straightforward, that I now turn.

\subsection{Spinfoams}

In spacetime dimension 2, ``quantum gravity" (whatever this means in 2d) can be defined in terms of the matrix models. In spacetime dimension 3, quantum gravity can be defined in terms of a beautiful construction due to Ponzano and Regge, that goes under the name of Ponzano Regge model \cite{Ponzano:1968uq}.  Boulatov \cite{Boulatov:1992vp} was able to show that the the matrix models can be generalized to 3d and the result is the Ponzano Regge model. In 1992, Ooguri \cite{Ooguri:1992eb} generalized Boulatov construction to four spacetime dimensions.  The resulting theory does not describe quantum gravity. It describes the quantum version of BF theory: the theory defined by the action
\be
      S_{BF}[B,\omega]=\int B\wedge F[\omega]
      \label{actBF}
\ee 
where $B$ is a two-form with values in the Lie algebra of a group $G$, $\omega$ is a $G$ connection and a trace in understood. Ooguri theory is formally defined over a two-complex by equation \eqref{int1} (with $SU(2)$ replaced by $G$) and
\be
A_{v}(h_l)=\int_{G} dg'_e \prod_l \delta(h_l,g_{s_l}g^{-1}_{t_l})
\label{vabf}
\ee
instead of \eqref{va}. BF theory is topological, in the sense that it has no local degrees of freedom, like general relativity in 3d, but unlikely general relativity in 4d, which has local degrees of freedom (such as gravitational waves). 

A surprising aspect of Ooguri's theory is that its states are spin networks, and its dynamics can be viewed as acting on the nodes of the spin networks. A very similar structure to loop quantum gravity, indeed.  

Suppose we wanted to exponentiate the loop-quantum-gravity hamiltonian evolution, as Feynman did in his first construction of the path integral, and give loop quantum gravity a ``sum-over-path" formulation. The general covariant evolution of a graph, where dynamics acts on nodes is naturally represented by a two-complex. It was therefore natural to explore the possibility to write such a ``sum-over-path" formulation of quantum gravity as a sum defined over two-complexes.  The idea of transforming loop gravity into a sum over foams was first proposed by Michael Reisenberger, and published in \cite{Reisenberger:1996pu} and, independently by Junichi Iwasaki \cite{Iwasaki:1995vg}.  

This possibility was particularly intriguing in view of Ooguri's result that such a structure was precisely what was needed to define quantum BF theory. And Ooguri theory was yes a topological theory, but also the only known example of a generally covariant quantum theory in four dimensions.  A research program then was open: find the appropriate variant of Ooguri's theory that would describe general relativity, namely a non-topological theory. 

The task did not seem too hard at the beginning because general relativity and BF theory are close relatives. The action of general relativity can be written in the form.
\be
      S_{GR}[e,\omega]=\int B\wedge F[\omega]
      \label{GRa}
\ee 
(very much like \eqref{actBF}), where the group is $SL(2,\mathbb{C})$ and
\be
     B:=(e\wedge e)^*.
     \label{simplicity}
\ee
The star indicates the Hodge dual in the $SL(2,\mathbb{C})$ algebra (that is, contracting two antisymmetric Minkowski indices with $\epsilon_{IJKL}$). In Ooguri's theory, the BF field is immediately integrated away, but it reappears as the variable conjugate to the connection, namely the generator of the $G$ action on the states.  To reduce the Ooguri theory to gravity, the task was simply to restrict each intermediate sum over states to the states where the $SL(2,\mathbb{C})$ generators satisfy  \eqref{simplicity}. Solving this problem took ten years. 

A first tentative solution was found by Barret and Crane with a celebrated model that goes under their names \cite{Barrett:1997gw}. The Barrett-Crane model is a non trivial modification of Ooguri's theory where equation \eqref{simplicity} is implemented \cite{DePietri:1999bx}.  However, evidence grew rapidly that the Barrett-Crane implementation of  \eqref{simplicity} was too strong: the resulting theory has fewer degrees of freedom than general relativity. The nail on the coffin of the old Barrett-Crane model was the first successful calculation of the propagator of the graviton from the theory: its diagonal components turn out to be correct, but its non-diagonal one turned out to be wrong \cite{Alesci:2007tx}. The correct (as far as we know today) solution appeared simultaneously in 2007 in the work of different groups \cite{Engle:2007uq,Engle:2007qf,Engle:2007wy,Perez:2003uq,Freidel:2007py,Livine:2007vk,Kaminski:2009fm}, and is in fact a quite simple variant of the Barrett-Crane model. What is needed to implement it is first to modify the general relativity action (without changing its equations of motion) adding a parity violating term \cite{Holst:1996fk}
\be
      S_{GR}[e,\omega]=\int B\wedge (F[\omega]+\frac1\gamma F[\omega]^*)
\ee 
This is the same step as the addition of the $\theta_{QCD}$ term in QCD:
\be
      S_{QCD}[A]=\int F[A]^*\wedge (F[A]+\theta_{QCD} F[A]^*)
\ee 
which also does not affect the classical equations of motion but affects the quantum theory.  The deep physical reason loop quantum gravity does not work without this extra term is not clear to me.  The constant $\gamma$ is called the Barbero-Immirzi parameter \cite{Barbero:1993aa,Immirzi:1996di}. Adding this extra term is the same as using the action  \eqref{GRa} but replacing \eqref{simplicity} by 
\be
     B:=(e\wedge e)^* + \frac1\gamma(e\wedge e)
          \label{simplicity2}
\ee
Now consider the restriction of the field to a 3d boundary. Choose a time-gauge defined by (I follow here \cite{Wieland:2010ec})
\be
        ne=0
\ee
where a scalar $n$ on the boundary, with values in Minkowski space ($ne=n_Ie^I$) has been chosen. Notice that the ``electric" $K=nB$ and ``magnetic" $L=-nB^*$ components of $B$ with respect to the time gauge\footnote{The decomposition of $B$ into $(K,L)$ is the same as the decomposition of the Maxwell field $F_{\mu\nu}$ into electric and magnetic field $(\vec E,\vec B)$.} satisfy 
\be
        K=\gamma L.
\label{si}
\ee
In the quantum theory, $K$ and $L$ become the generators of boosts and rotations of $SL(2,\mathbb{C})$. The problem of reducing Ooguri theory to quantum GR is therefore the problem to see if there is a set of $SL(2,\mathbb{C})$ states where boost and rotation generators are related by \eqref{si}. The answer is positive, and the kernel $K$ defined in \eqref{K} implements precisely this projection  \cite{Wieland:2010ec,Ding:2010ye,Ding:2010fw}. This gives the dynamics defined in sect \ref{dyn}.

\subsection{Group field theory}

The peculiar feature of the definition of 2d quantum gravity in terms of matrix models is that the sum over 2d geometries is generated as the Feynman expansion of an auxiliary quantum theory ---in fact, a quantum theory of matrices.  The same happens for Boulatov theory, where the amplitudes of the Ponzano-Regge model are recovered as Feynman amplitudes of an auxiliary field theory. In this case the auxiliary theory is a peculiar nonlocal quantum field theory for a field defined on a group manifold.  The same is true for Ooguri's theory in 4 dimensions. These auxiliary field theories are called ``group-field-theories".  It turns out that the Barrett-Crane model can also be expressed a group field theory \cite{DePietri:1999bx}, and a group field theory for the dynamics of loop quantum gravity can be constructed and is currently under investigation \cite{Oriti06}. 

The advantage of the group-field-theory formulation are technical as well as conceptual.  On the technical side, the amplitudes \eqref{int1} of quantum gravity can be interpreted as rather conventional Feynman graphs and this brings large parts of the full machinery of standard quantum field theory back into play.  In particular, an ongoing effort is course for understanding scaling, renormalization and potential divergences of the theory, using this language \cite{Geloun:2010vj,Krajewski:2010yq,Rivasseau:2010kf}. On the conceptual side, the construction sheds some light on the double aspect that the loop quantum gravity dynamics: foams can be seen both as Feynman graphs and as spacetime lattices. This is precisely the same as what happens in 2d gravity matrix models. 

\subsection{Loop gravity as a generalized TQFT}

There is another useful perspective on the theory defined in Section \ref{art}, which clarifies the sense in which it is a generally covariant quantum field theory.  

A general framework for defining generally-covariant quantum field theory is provided by the definition of topological quantum field theory (TQFT) proposed by Atiyah \cite{Atiyah:1988fk,Atiyah:1990uq}.  In Atiyah's scheme, an $(n+1)$-dimensional TQFT is as a functorial association of a finite dimensional Hilbert  space $H_\Sigma$ to each closed oriented $n$-manifold $\Sigma$, and a vector $Z_M\in H_{\partial M}$ to each oriented $(n+1)$-manifold $M$. Here, a spinfoam model associates a Hilbert space $H_\Gamma$ to each oriented graph $\Gamma$, and a vector $Z_C\in H_{\partial{\cal C}}$, given by \eqref{int1}, to each foam $\cal C$ with boundary $\Gamma$. This association satisfies the axioms
\begin{itemize}
\item
(multiplicativity) $$H_{\Gamma_1\cup\Gamma_2}=H_{\Gamma_1}\otimes H_{\Gamma_2}$$
\item
(duality) $$H_{\overline{\Gamma}}=H_{\Gamma}^{*}\quad\textrm{and}\quad Z_{\overline{\C}}=Z_{C}^{\dagger}$$
\item
(functoriality) $$Z_{\C_1\cup_{\Gamma}\C_2}=\langle Z_{\overline{\C_2}}\vert Z_{\C_1}\rangle_{H_{\Gamma}}=\langle Z_{\overline{\C_1}}\vert Z_{\C_2}\rangle_{H_{\overline{\Gamma}}}$$
\end{itemize}
where the bar over graphs and foams indicate reversing links and edges orientation \cite{Rovelli:2010qx}. This set of axioms is precisely Atiyah's, except for two points.\footnote{See also Oeckl's \emph{general boundary} formulation of quantum field theory \cite{Oeckl:2005bv}.} First, because the boundary graphs are not oriented globally, like boundary manifolds, but only linkwise, the condition of functoriality cannot be framed in terms of morphisms between source and target spaces, but rather in terms of partial inner products as above. Second, we do \emph{not} require the Hilbert spaces to be finite dimensional.

Therefore loop gravity is essentially a TQFT in the sense of Atiyah, where the cobordism between 3 and 4d manifold is replaced by the cobordism between graphs and foams.  What is the sense of this replacement?  

TQFT defined on manifolds are in general theories that have no local degrees of freedom, such as BF or Chern-Simon theory, where the connection is locally flat. Its only degrees of freedom are global ones, captured by the holonomy of the connection wrapping around non-contractible loops in the manifold.   In general relativity, we do not want a flat connection: curvature is gravity.  But recall that the theory admits truncations \`a la Regge where curvature is concentrated in $d-2$ dimensional submanifolds. If we excise these $d-2$ submanifolds from the Regge manifold, we obtain manifolds \emph{with $d-2$ dimensional defects}. The spin connection on these manifolds is locally flat, {\em but it is still sufficient to describe the geometry, via its non trivial holonomies wrapping around the defects}  \cite{Bianchi:2009tj}. In other words, general relativity is approximated arbitrarily well by a connection theory of a \emph{flat} connection on a manifold with (Regge like) defects. Now, the relevant topology of a 3d manifold with 1d defects is precisely characterized by a graph, and the relevant topology of a 4d manifold with 2d defects is precisely characterized by a two-complex. In the first case, the graph is the 1-skeleton of the cellular complex dual to the Regge cellular decomposition. It is easy to see that this graph and the Regge manifold with defects have the same fundamental group.  In the second case, the two-complex is the 2-skeleton of the cellular complex dual to the 4d Regge cellular decomposition. In this case, the faces of the two-complex wrap around the 2d Regge defects. Therefore equipping Atiyah's manifolds with  $d-2$ defects amounts precisely to allowing local curvature, and hence obtaining genuinely local (but still generally covariant) bulk degrees of freedom.

\section{Applications and results}\label{applications}

What follows is not an extensive illustration of all tentative applications of the theory, nor does it reflect the amount of published literature on the different topics.  I only focus on a few directions that in my opinion have led to particularly relevant physical results so far. 

\subsection{Loop cosmology: canonical}

The most spectacular and successful physical application of loop gravity, in my opinion, is loop cosmology. The physical problem is that at early cosmic times the universe was not in the regime where classical general relativity holds, and quantum gravitational effects presumably dominate.  Started by Martin Bojowald and largely developed by Abhay Ashtekar, Parampreet Singh, Alejandro Corichi and their collaborators, loop cosmology is an application of the basic ideas of loop gravity to the description of the physics of the universe at very early time \cite{Bojowald:2008zzb,Ashtekar:2008zu}. The theory is constructed following the conventional idea of quantizing cosmological models, but using crucial physical inputs from loop gravity in the construction. Physically, the key ingredient is the quantization of the geometry at small scale, and the existence of the area gap \eqref{gap2}. Intuitively, this means that at very small values of the scale factor, the universe enters in a region where the discreteness of the geometry becomes the dominant physical effect. The cosmological Wheeler-deWitt equation can be written in this language, and it turns out to be a finite difference equation instead of a differential equation.  Numerical analysis and analytic approximations can then be used to study the early cosmological evolution. The result, for a Robertson-Walker metric is a correction of the Friedmann dynamics by quantum effects. The Friedmann equation 
\be
      \left(\frac{\dot a}a\right)^2=\frac{8\pi G}3\rho-\frac{k}{a^2}+\frac{\Lambda}{3}
\ee
gets corrected by the factor 
\be
      \left(\frac{\dot a}a\right)^2=\frac{8\pi G}3\rho\left(1-\frac{\rho}{\rho_c} \right)-\frac{k}{a^2}+\frac{\Lambda}{3}
\ee
where the critical density is 
\be
\rho_c = \left(\frac{8\pi G}3 \ \gamma^2 a_o\right)^{-1}.
\ee
In a collapsing universe ($\dot a<0$) when the matter energy density $\rho$ reaches this Planck scale value, $\dot a$ vanishes, and the universe stops contracting. The integration of the equation shows that the universe bounces back, and enters in an expanding phase. Numerical analysis of the full Wheeler-deWitt equation confirms that this is indeed the behavior of a semiclassical wave packet. In other words, quantum effects resolve the initial singularity in precisely the same manner in which the singularity of a pointlike-electron falling into the infinite Coulomb potential well of a nucleus is resolved by quantum mechanics. 

What happens is that quantum effects act as an effective quantum repulsion, that allows the universe to bounce back from a collapse. It is worth pointing out that this does not happen with conventional quantum cosmology models, namely with the old cosmological Wheeler-deWitt equation.  The bounce is truly an effect of the quantization of geometry. 

The result is obtained without fine tuning initial conditions, and without imposing a boundary condition at the singularity.  It is quite robust, in the sense that it happens in all cases analyzed so far. In particular, the result has been extended to open and closed universes, to anisotropic models, to different matter couplings, to cosmology with inhomogeneities, and to Gowdy cosmologies \cite{MenaMarugan:1997us,Corichi:2007ht,MartinBenito:2008ej,Garay:2010sk}  among others. See \cite{Ashtekar:2007em} and references therein.  

The same technique that leads to the resolution of the initial cosmological singularity has also been shown to lead to the resolution to all other (``strong") cosmological singularities \cite{Singh:2009mz,Singh:2010fk}, and even to the singularities inside black holes \cite{Modesto:2004xx,Ashtekar:2005qt,Gambini:2008dy}. The standard singularity theorems of Penrose, Hawking and others are not contradicted because these suppose the Einstein equations, while loop quantum effects modify these equations at small scale. 

The situation can be compared with the effect of the Fermi-Dirac statistics, associated with the quantum nature of matter, that can bring to equilibrium a stellar collapse, leading to stable white dwarfs and neutron stars. If the total mass of the star is larger than a critical value, classical gravity overwhelms this force and the star continues to collapse. Loop gravity indicates that there is a further repulsive force associated now with the quantum nature of geometry, and this is strong enough to counter the classical gravitational attraction, irrespective of how large the mass is.

There is much more in loop cosmology than what I can cover here. For instance, from the modified Friedmann equation and the standard conservation law it follows that the deceleration equation is modified to
\be
\frac{\ddot{a}}{a}=-\frac{4}{3}\pi G\left[\rho\left(1-4\frac{\rho}{\rho_c}\right) + 3P\! \left(1-2\frac{\rho}{\rho_c}\right) \right]+\frac{\Lambda}{3},
\ee
so that $\ddot a$ can be positive, opening the way to the onset of inflation. In addition, it has recently been shown that in the presence of an appropriate scalar field, inflation is generic in loop cosmology, and does not require unlikely initial conditions \cite{Mielczarek:2009fk,Ashtekar:2009mm}.

This does not mean of course that all is clear in the physics of the very early universe. First, so far the bounce has been derived in models with few degrees of freedom, or perturbations of the same. What happens, and what is the full nonperturbative quantum geometry at the bounce? Second, can we connect the bounce picture to the open puzzles concerning the low entropy of the initial state of the universe? And most importantly, in my opinion, questions are open concerning the physical interpretation of the bounce itself: does it make sense to say that  ``times continues" across a region where spacetime quantum fluctuations dominate? Calculations so far are in terms of the simple minded clock-time provided by the value of a scalar field.  In my opinion, an interpretation less tied to the idea of a \emph{continuation of physical time} across the quantum region is needed.  

Still, the resolution of the initial singularity is a remarkable success of loop gravity, which goes well beyond its initial hopes. I find extremely exciting that questions such as the ones above can be posed and studied, not in the context of free speculation, but within a theory that has no other physical assumptions than general relativity and quantum mechanics.

\subsection{Black hole entropy}

Another success of the theory, although perhaps not as complete, is the resolution of another long standing open problem in quantum gravity, namely the first principles computation of the entropy of a black hole,
\be
   S=\frac14 \frac{A}{\hbar G}
   \label{be}
   \ee
derived by Bekenstein and Hawking from semiclassical considerations. Loop gravity offers a precise scenario for understanding the physical origin of this entropy. I illustrate here this scenario in the way I understand it, a way that might perhaps not be in complete agreement with the views of several researchers working in this direction. 

The expression ``black hole" is used in the physics literature in ways which are sometimes a bit ambiguous, because it indicates both the stationary black holes (as when it is said that a black hole is uniquely characterized by its mass, charge and angular momentum) and non stationary black holes (which are definitely \emph{not} uniquely characterized by mass, charge and angular momentum). Strictly speaking, no realistic  black hole is stationary, since it interacts with a moving environment. Therefore, strictly speaking a physical black hole is not characterized by just its mass, charge and angular momentum, but by an infinite number of parameters, which can be understood as the amplitudes of a mode expansion of its geometry in multiple moments, or, more simply, as the variable ``shape" of its horizon.  This quickly becomes negligible in astrophysics, because a massive black hole rapidly radiates away its multiple moments and is not much affected by its surroundings unless a substantially massive body is in its vicinity.  But the picture of a black hole that decays rapidly to a stationary solution becomes incorrect in two situations: in a thermal situation in which the hole is in touch with a thermal bath; and where we do not disregard quantum effects, since the horizon modes may keep ``zero-point energy" quantum fluctuations. A fortiori, a black hole cannot be assumed to be stationary if we are interested in the way quantum physics affects its thermal behavior. 

This means that to describe the effects of quantum physics on the thermal behavior of the hole we must take into account the fluctuations of its horizon.  Notice that the \emph{interior} of a black hole is of no interest here because it cannot affect the way the hole interacts thermally with its surrounding (which is what is of interest), being causally disconnected from it. And the \emph{exterior} of the hole is of no interest either, since its eventual entropy can be described by appropriate standard quantum field theoretical methods. Therefore what is of interest for black hole entropy can only be the set of degrees of freedom on the horizon itself. These exist because the black holes that are thermal cannot be stationary. 

More precisely, we should use the standard distinction between the macroscopic state of a system and the microscopic state.  While we are interested in macroscopic states that are stationary, this obviously does not mean the microscopic states must be so as well. A  gas with constant uniform density and pressure in a box is in a stationary state, but this does not imply that its molecules stand still!

So what are the "molecules that move", in the case of a black hole? The above discussion makes it clear: they are given by the variable shape of its horizon. Here ``shape" may refer to both its intrinsic and its extrinsic geometry, since both interact with the surroundings. Thus, even fixing the intrinsic geometry of the horizon itself, there remains all the infinite degrees of freedom that describe the way this geometry sits in spacetime.  The entropy of a black hole can then be computed, as for any physical system, as the logarithm of the number of the physical states of this system, compatible with a given macroscopic state, for instance fixing area, charge and angular momentum of a macroscopic stationary black hole. In the classical theory, this number is infinite, because the horizon is a continuous surface.  In loop quantum gravity, this number is finite, because of the quantization of the geometry. 

The actual counting of the degrees of freedom on the horizon is a delicate task, which exploits the details of the quantum state space of loop gravity, and I will not enter into the calculations here. See the standard reference \cite{Ashtekar:2000eq} or a recent review such as \cite{Ashtekar:1999ex} for a detailed calculation and more complete references.  

The discussion on further clarifications and ameliorations of the derivation is still ongoing. See in particular \cite{Engle:2010kt}, where a number of issues in previous derivations have been cleaned up and clarified.  The results of these calculations are that the entropy is finite, determined by the quantization of the geometry and given by 
\be 
   S_{loop}=\frac14 \ \frac\gamma{\gamma_o}  \frac{A}{\hbar G}
   \label{bel}
\ee
where $\gamma_o$ is a number of the order of unity that results from the combinatorial calculation. Its most updated value \cite{Meissner:2004ju,Ghosh:2004wq,Agullo:2009eq}, as far as I know, is $\gamma_o=0.274067...$ Thus, the entropy computed in loop quantum gravity, namely $S_{loop}$ is equal to the Bekenstein-Hawking entropy \eqref{be} if the value of the Barbero-Immirzi parameter is
\be
       \gamma = \gamma_o= 0.274067... \
       \label{gamma}
\ee
This result is then often taken as a way to compute the value of $\gamma$, which is fixed to this value and used as such in other contexts.

Remarkably, the calculation can be repeated for different kind of (macroscopic) black holes and this \emph{same} value of $\gamma$ gives the correct Bekenstein-Entropy in all cases (contrary to claim of the opposite that for reasons that I do not understand can be found on the internet.) The fact that the same value of $\gamma$ works in all case is a strong test of consistency of the entire derivation. 

Another recent fresh approach to the same problem, compatible with the one mentioned and yielding to the same result, is in \cite{Bianchi:2010qd}, 
where the loop input is used a bit differently and the counting of microstates is done via a mapping to an equivalent statistical mechanical problem: the counting of conformations of a closed polymer chain. The correspondence suggests a number of intriguing relations between the thermodynamics of black holes and the physics of polymers.

I think that these black hole entropy results can be counted as a success of loop gravity for various reasons.  First, the entropy turns out to be finite, to be proportional to the area, to be of the right order of magnitude, and a clear and compelling physical picture of the origin of this entropy exists. 

However, the result is not entirely satisfactory in my opinion.  It is not strange that a fundamental parameter in the theory could have a peculiar value such as $\gamma_o$: we do not understand the origin of other fundamental constants, such that the fine-structure constant. But it \emph{is} strange, and perplexing, that there be such a peculiar parameter in the theory, which then cancels exactly with a number that characterizes a complicated statistical counting, in such a way to give a round number such as 4 in \eqref{be}.\footnote{In addition, as pointed out by Ted Jacobson \cite{Jacobson:2007uj}, radiative correction could play a role, affecting the Area observable and the coupling constants. In particular, the $G$ entering \eqref{be} is the Newton constant at large distance, while the $G$ entering \eqref{bel} might be the Newton constant $G_{Plank}$ at the Planck scale. So that we should rather pose
\be
       \gamma = \gamma_o\frac{G_{Plank}}{G}.
\ee
and the numerical value of $\gamma_o$ could perhaps reappear in the relation between Plank scale and infrared physics.}  I think that the sense that there is something important which is not yet understood is unavoidable.

\subsection{$n$-point functions}

A formalism for computing $n$-points functions has been put into place in recent years and this can be counted as an essential part of the theory, in my opinion. 
The computation of graviton $n$-point functions from the background-independent formalism of section \ref{art} is of interest for various reasons. 

First, because the traditional difficulty of background-independent quantum gravity is to make contact with observations and standard physics.  The problem is seriously felt in the canonical version of the theory, in spite of a number of courageous and determined recent attempts to address it. Quantum gravity is structurally different from conventional quantum field theory: field operators do not carry a position label $x$, or momentum $p$, that allow us to use standard methods to go from the theory to actual measurements, say, for instance, of scattering amplitudes. 

Second, computing $n$-point functions (in the low energy limit) is in principle a complete way to test whether the classical limit of the theory correctly gives general relativity. This ``test" role of the $n$-point functions calculation has already proven very powerful, by showing that the ``old" Barret-Crane model was incomplete as a quantum theory of general relativity. 

Finally, if we could compute $n$-point functions systematically, we would be able to compute all quantum corrections to classical GR in appropriate perturbative expansions, and, at least in principle, resolve the problem of the non-renormalizability of general relativity, by explicitly computing from first principles the values of the infinite number of renormalization parameters. 

We are not yet there and the calculations of $n$-point functions are still difficult within the theory. Still, a general procedure for constructing them is in place, and several partial results have been obtained. Let me illustrate both these points. 

An $n$-point is a quantity defined over a background, such as Minkowski space, because it expresses correlations between fluctuations over this background.  Therefore we should provide the background-independent theory with an input about the background, in order to define such functions.  After some early attempts to introduce the background directly in the bulk \cite{Iwasaki:1992qy}, it was later understood that the proper way to introduce the background is via the  boundary formalism. The idea behind this formalism is to consider semiclassical coherent states on finite boundaries of 4d regions, and compare the amplitude of such states with the amplitude of the same state over which field excitations are added \cite{Oeckl:2003vu,Rovelli:2004fk}.

To do so, the first ingredient needed is a way to deal with semiclassical states. As mentioned in sect. \ref{spin}, this has been developed. The second ingredient is the large distance expansion of the vertex amplitude \eqref{va}. This is available as well \cite{Barrett:2009mw}. The third ingredient is an understanding of the appropriate expansion needed. Obviously we cannot hope for exact calculations in the theory and a perturbation expansion of some sort is needed.  

The theory itself provides three natural expansion parameters. The first is given by the truncation provided by cutting the theory to a finite $\Gamma$. This is analogous to truncating to a finite number of particles, as we do in QED.  The second, related, expansion in the two-complex. Again, this corresponds to fixing the order of the Feynman graphs in a QED perturbative calculation. Finally, a large $j$ expansion is natural in the theory.  Large $j$ means large distances, or, equivalently, large quantum numbers. In both cases, in quantum gravity this means to go away from the quantum regime, and go towards the classical theory. Quantum corrections can therefore be organized in $1/j$. As for any expansion, these expansion will be good in \emph{some} regime, but for the moment, a clean characterization of their regime of validity is missing, and exploration is by attempts. 

Using all this, the calculation of the 2 point function of the theory at the lowest order in these expansions has been completed, in the context of the euclidean version of the theory ($SL(2,\mathbb{C})$ replaced by $SO(4)$). The remarkable result is the recovery of the two-point function of general relativity.  The test that was failed by the old Barrett-Crane model is then passed by the improved version of the model, which is the one currently studied, and defined by equations (\ref{int1},\ref{va},\ref{K}) above.

	\subsection{Spinfoam cosmology}

Loop quantum cosmology uses the canonical formalism. Can we treat cosmology directly in term of the covariant spinfoam dynamics?  The answer turns out to be positive and offers an independent element in support to the conjecture that the theory defined in Section \ref{art} yields general relativity in the classical limit.  By using the coherent state technology and the expansions described in the previous paragraph, it is possible to compute the evolution of a homogenous isotropic coherent state of the universe.  The calculation can be performed \cite{Bianchi:2010zs}, and the result appear to be consistent with the Friedmann evolution. In particular, if a cosmological constant term is added to the quantum theory, the de Sitter solution of the Einstein equations emerges from the full quantum theory \cite{Bianchi:2011ym}.
 
The relation between this spinfoam formulations of loop cosmology and the canonical one is under investigation \cite{Ashtekar:2009dn,Ashtekar:2010ve,Campiglia:2010jw,Henderson:2010qd,Calcagni:2010ad} and is likely to shed new light on the overall structure of the theory.
\vskip .5cm
In closure of this chapter, allow me to repeat that this is only a short overview and I am forced to leave out numerous other interesting developments.

\section{What has been achieved?}\label{assessment}

Finally, let me try a tentative evaluation of the state of the art, and a comparison with the initial hopes raised by the theory.  I start from the negative elements

\subsection{Open problems}

\begin{description}

\item[\em Experiments and predictions.] Needless to say, there are no experiments supporting this (or any other) quantum theory of gravity. All current theories of quantum gravity are in the realm of the theoretical attempts.  

But the situation is more serious than just this. Loop gravity, as well as all other quantum theories of gravity, has so far been incapable of producing a single 
clear-cut prediction that could in principle put the theory under cogent empirical test. This is bad and is a weakness of today's fundamental theoretical physics.  

We need definite predictions, like those that all the good physical theories of the past have been able to produce.  A theory that is compatible with any possible future measurement outcome is not a serious theory. A theory that predicts everything and the opposite of everything is not a mine of ideas: it is a bucket of confusion.

To be sure, loop gravity {\em makes} definite predictions.  For instance, any measured physical area (such as any total scattering amplitude) must turn out to be quantized, and given by the area spectrum \eqref{spectrum} computed within the theory. For an intriguing recent suggestion on the effect of angle quantization on  scattering, see \cite{Major:2010qg}.\footnote{Interesting speculative ideas are also explored, such as the possibility that small stable black holes are compatible with the theory and could be a component of dark matter, emitting ultra-high-energy-cosmic-rays  compatible with the observed rate of the Auger detector \cite{Modesto:2009fq}.}  But these predictions are still far from present measurements, as far as I can see. 

It is important to notice that the Planck regime is not unreachable by present technology \cite{AmelinoCamelia:2008qg}, as was believed to be not long ago. For instance, proton decay experiments reach phenomena at energies not too far from the Planck scale. Gamma-ray burst time-delay measurements put limits to dispersion-relations modifications that reach the Planck scale. Analyses such as the recent  one of the Crab nebula test energies well beyond the Planck scale  
\cite{Jacobson:2005bg}. But so far, no overlap has been found between what the theory predicts and what we are capable to measure.  

Recent Planck-scale observations appear to support local Lorentz invariance \cite{Jacobson:2005bg}. This has been erroneously presented by some authors as evidence against loop gravity.  I  want to stress the fact that loop gravity does {\em not} imply a violation of Lorentz invariance. In particular, the naive argument, often heard, that a minimal length is incompatible with Lorentz invariance is wrong, because it disregards quantum theory. The same argument would imply that a minimum value for a component of the angular momentum would be incompatible with rotation invariance, a conclusion manifestly contradicted by the quantum mechanics of angular momentum. For a complete discussion of this point, see \cite{Rovelli:2002vp}. The Lorentz invariance of the loop and spinfoam formalism can be made manifest%
\footnote{Using the Livine-Alexandrov ``projective spin network" \cite{Livine:2002ak,Alexandrov:2002br,Alexandrov:2010pg} technique.}: see 
 \cite{Rovelli:2010ed} and references therein.   For the moment, existing theoretical evidence is against Lorentz violations \cite{Collins:2004bp}, and in accord with observations. So, for now we have no useful information from this direction.

The most likely window of opportunity at present seems to come from early cosmology.  Quantum effects could be significative shortly before the onset of inflation and could affect, for instance, the CMB at multiple moments somewhat higher than the ones presently measured.  The hope that the theory could provide an input to early cosmology sufficient for predicting observable quantum effects, and interesting attempts in this direction exist \cite{Grain:2009kw,Barrau:2010nd,Barrau:2009fz,Grain:2010yv,Mielczarek:2010bh,Bojowald:2010zz}. But for the moment, I see no definite prediction that could be used to falsify the theory. To make loop (or any other) quantum theory of gravity, physically credible, this must change. 

\item[\em Divergences.] The theory has no ultraviolet divergences. This can be shown
in various ways, for instance rewriting \eqref{int1} in the spin-network basis and noticing that the area gap makes all sums finite in the direction of the ultraviolet.  However, divergences might be lurking elsewhere, and they probably are. There might indeed be infrared divergences, that come from large $j$.  The geometrical interpretation of these divergences is known. They corresponds to the ``spikes" in Regge calculus: imagine taking a triangulation of a two-surface, and moving a single vertex of the triangulation at large distance from the rest of the surface. Then there will be a cycle of triangles which are very lengthened, and have arbitrary large area.  This is a spike. 

A number of strategies can be considered to control these infrared divergences. One is to regularize them by replacing groups with quantum groups. This step has a physical ground since this modification of the vertex amplitude corresponds to adding a cosmological constant to the dynamics of the theory. The theory with the quantum group is finite \cite{Fairbairn:2010cp,Han:2010pz}.  

The second possible strategy is to see if infrared divergences can be renormalized away, namely absorbed into a redefinition of the vertex amplitude. A research line is active in this direction \cite{Krajewski:2010yq,Geloun:2010vj}, which exploits the group-field-theory formulation of the theory.\footnote{Infrared divergences in spacetime appear as ultraviolet divergences on the group manifold.} Finally, the third possibility is that infrared divergences could be relatively harmless on local observables, as they are in standard field theory.  

\item[\em Canonical versus covariant.] The kinematics of the canonical theory and the covariant theory are identical. The dynamics defined in the two versions of the theory ``resemble" one another, but it is has not yet been possible to clearly derive the relation in the 4d theory (in the 3d theory, the corresponding relation has been nicely proven \cite{Noui:2004iy}. See also, in the cosmological context \cite{Ashtekar:2009dn,Ashtekar:2010ve,Campiglia:2010jw,Henderson:2010qd}.) This is another form of incompleteness of the theory. 

\item[\em Classical limit.] There is substantial circumstantial evidence that the large distance limit of the theory is correctly general relativity, from asymptotic analysis and from large distance calculations of $n$-point functions and in spinfoam cosmology; 
and there are open directions of investigations to reinforce this evidence. The degrees of freedom are correct and the theory is generally covariant: the low-energy limit is not likely to be much else than general relativity.  But there is no solid proof yet. 

\item[\em Incompleteness.] Finally, there is truly very much that is still missing: computing $n$-functions to higher order and for larger $n$ and comparing with general relativity, and much more. The theory is far from being complete. 

\end{description}

\subsection{Achievements}

Up to the limitations mentioned above, I think it is safe to say that loop quantum gravity has fulfilled and in fact largely outperformed the initial hopes it has raised. The theory has encountered a number of technical and conceptual stumbling blocks during its development (over-completeness of the loop basis, difficulties of extracting physics from the canonical theory, defining a finite Wheeler-deWitt constraint, finding a way to compute background-dependent quantities from the background independent theory, correcting the Barrett-Crane model, and many others) but these have all been nicely  circumvented or overcame.  

Major surprises have appeared from the theory itself.  Loop quantum gravity had not been initially conceived to describe quantized areas and volumes. The discreteness of the geometry has emerged from solving the spectral problem of the corresponding operators.  The convergence between the kinematic obtained from a canonical quantization of general relativity and that obtained from the modification of the covariant Ooguri theory to reduce BF to gravity, has been unexpected and surprising. The resolution of the Big Bang singularity was not expected when the theory was first applied to cosmology, and so on. 

In my opinion, the theory provides today a tentative, but intriguing possible formalism for describing quantum spacetime, making predictions at the Planck scale, dealing with the early universe, and offering a coherent conceptual framework where general relativity and quantum mechanics cohabit amicably. 

What I find particularly attractive is that the theory is based on no additional physical hypothesis other than general relativity and quantum mechanics, two theories that have today endless empirical backing.  In fact, the theory is based on taking seriously these theories, and pushing them beyond the regimes where they were conceived.  This is a strategy that has often been effective in the history of physics. 

And I find the resulting physical picture compelling: space made up by quanta of space, whose dynamics is given by a simple expression like equations (\ref{int1},\ref{va},\ref{K}). The double interpretation that merges the formal structures used in perturbative QED (Feynman graphs) and nonperturbative QCD (lattice discretization) is also particularly compelling: two distinct characteristic technologies of quantum field theory turn out to be more strictly related than expected, when taking into account the basic discovery of general relativity: that spacetime is nothing else than a physical field like any other. The lattice sites where the QCD field are defined can be viewed  themselves as quanta of another field: the gravitational field. 

On the other hand, I think that the theory is still fragile and I am far from sure that it is physically correct or that it has already reached its definite form.  Its classical limit might finally turn out to be wrong, the calculation of higher $n$-point functions could turn out to be inconsistent or impossible. And even if the theory was fully coherent, of course, nature could have chosen something else.  We'll see. But the situation in quantum gravity is in my opinion by very far better than twenty-five years ago, and, one day out of two, I am optimistic. 

\vskip1cm\centerline{-----}\vskip1cm

Thanks to the CQG referee for numerous valuable suggestions, and particular thanks to Leonard Cottrell for his careful reading of the manuscript.


\end{document}